\documentclass[useAMS,usenatbib]{mn2e}

\usepackage{graphicx,latexsym}
\usepackage{natbib}
\bibpunct{(}{)}{,}{a}{}{,}
\usepackage{rotating,ams}
\usepackage[ps2pdf,linktocpage]{hyperref}
\hypersetup{%
 colorlinks=false,
 bookmarksopen=true,
 bookmarksnumbered=true,
 pdfstartpage={1}, 
 pdftitle = {The Antennae star cluster system revisited I: Luminosity functions},
 pdfauthor = {P. Anders, N. Bissantz, L. Boysen, R. de Grijs, U. Fritze - v. Alvensleben} ,
 pdfkeywords = {extragalactic astrophysics, young star clusters, Antennae galaxies, evolutionary
 synthesis modelling, GALEV, luminosity function, completeness function},
 pdfproducer = {LaTeX with hyperref}
}

\defcitealias{2006A&A...451..375A}{AC/sizes Paper}
\def\spose#1{\hbox to 0pt{#1\hss}}
\def\lta{\mathrel{\spose{\lower 3pt\hbox{$\mathchar"218$}}
 \raise 2.0pt\hbox{$\mathchar"13C$}}}
\def\gta{\mathrel{\spose{\lower 3pt\hbox{$\mathchar"218$}}
 \raise 2.0pt\hbox{$\mathchar"13E$}}}

\begin{document} 

\title[The Antennae star cluster system revisited - The LF]{
The Young Star Cluster System in the Antennae: Evidence for a Turnover in
the luminosity function}

\author[P. Anders et al.]{P. Anders$^{1,2}$\thanks{E-mail:
P.Anders@astro.uu.nl}, N. Bissantz$^{3,4}$, L. Boysen$^4$, R. de
Grijs$^5$, and U. Fritze -- v. Alvensleben$^{1,6}$\\
$^1$ Institut f\"ur Astrophysik, University of G\"ottingen,
Friedrich-Hund-Platz 1, 37077 G\"ottingen, Germany\\
$^2$ Sterrenkundig Instituut, Universiteit Utrecht, P.O. Box 80000,
3508 TA Utrecht, The Netherlands\\
$^3$ Institut f\"ur Mathematische Stochastik, University of
G\"ottingen, Maschm\"uhlenweg 8-10, 37073 G\"ottingen, Germany\\
$^4$ Fakult\"at f\"ur Mathematik, Ruhr-University of Bochum, Mathematik III, NA 3/70,
Universit\"atsstra{\ss}e 150, 44780 Bochum, Germany\\
$^5$ Department of Physics \& Astronomy, The University of Sheffield,
Hicks Building, Hounsfield Road, Sheffield S3 7RH\\
$^6$ Centre for Astrophysics Research, University of Hertfordshire,
College Lane, Hatfield AL10 9AB \\}

{\date{Accepted ---. Received ---; in original form ---.}
%%\pubyear{2006}
%\begin{document}
\maketitle

\begin{abstract} 

The luminosity functions (LFs) of star cluster systems (i.e. the
number of clusters per luminosity interval) are vital diagnostics to
probe the conditions of star cluster formation. Early studies have
revealed a clear dichotomy between old globular clusters and young
clusters, with the former characterised by Gaussian-shaped LFs, and
the latter following a power law. Recently, this view was challenged
by studies of galaxy merger remnants and post-starburst galaxies. In
this paper we re-evaluate the young ($\lta$ few hundreds of Myrs, with
the majority $\lta$ few tens of Myrs) star cluster system in the
ongoing spiral-spiral major merger system NGC 4038/39, the
``Antennae'' galaxies. The Antennae galaxies represent a very active
and complex star-forming environment, which hampers cluster selection
and photometry as well as the determination of observational
completeness fractions. A main issue of concern is the large number of
bright young stars contained in most earlier studies, which we
carefully exclude from our cluster sample by accurately determining
the source sizes. The resulting LFs are fitted both with Gaussian and
with power-law distributions, taking into account both the
observational completeness fractions and photometric errors, and
compared using a likelihood ratio test. The likelihood ratio results
are rigidly evaluated using Monte Carlo simulations. We perform a
number of additional tests, e.g. with subsets of the total sample, all
confirming our main result: that a Gaussian distribution fits the
observed LFs of clusters in this preferentially very young cluster
system significantly better than a power-law distribution, at a 
(statistical) error probability of less than $0.5$ per cent. 

\end{abstract}


\section{Introduction}
\label{sec:intro}

\subsection{Star Clusters: evolution and interpretation}

Star clusters (SCs) form nearly instantaneously through the collapse
of giant molecular gas clouds (GMCs). Hence all stars within a SC are
approximately coeval, and represent a simple stellar population (SSP).
A small number of parameters, in particular their initial chemical
composition and initial stellar mass function (IMF), are enough to
describe their colour and luminosity evolution on the basis of a given
set of stellar evolutionary tracks or isochrones (e.g.
\citealt{2002A&A...392....1S,2003A&A...401.1063A,2003MNRAS.344.1000B}).
Therefore, observed spectrophotometric properties of SCs are
relatively easy and straightforward to interpret.

SC formation is a major mode of all star formation, and possibly even
the dominant mode in strong starbursts triggered in gas-rich galaxy
mergers (e.g., \citealt{1995Natur.375..742M,2003NewA....8..155D}).

SCs are excellent tracers of their parent galaxy's star-forming
properties. Radial age gradients in SC systems and age differences
between SF regions tell us about the dynamical evolution of a
starburst (\citealt{2001AJ....121..768D}). This is owing to the fact
that SCs are not only relatively easy to model but also much easier to
analyse than the integrated light of the galaxy, because SCs can be
studied individually. Even without individual SC spectroscopy,
multi-band imaging in at least four suitable passbands allows us to
determine the age, metallicity, extinction, and mass of a SC
(\citealt{2004MNRAS.347..196A,2004MNRAS.347...17A}). The age and
metallicity distributions of a SC system directly reveal the SF and
chemical enrichment histories of its parent galaxy, to much higher
precision than studies of the galaxy's integrated light, since the
latter is always dominated by the last major epoch of SF (see e.g.
\citealt{2004A&A...414..515F}). The long-lived SCs, and the old
globular clusters (GCs) in particular, hold a key role in this respect
(cf. \citealt{2004Natur.427...31W}).

\subsection{Star Cluster Systems}

The most commonly used diagnostics to explore the properties and
evolution of entire SC {\sl systems} are their luminosity and mass
functions\footnote{Following \citet{1996ApJ...457..578M} and
\cite{parmentier06} we denote a luminosity {\sl spectrum} (LS) the
number of objects per {\sl linear} luminosity interval ${\rm d}N/{\rm
d}L$, while we refer to the luminosity {\sl function} (LF) to describe
the number of objects per {\sl logarithmic} interval ${\rm d}N/{\rm
d}\log L$. Magnitudes (mag) are logarithmically related to
luminosities (L) $\log L \propto mag$. If a LS can be described by a
power law $L^{-\alpha_{LS}}$, the slope of the corresponding LF is 
$\alpha_{LF}=\alpha_{LS}-1$ and if expressed in magnitudes
$\alpha_{LF'}=0.4 \times (\alpha_{LS}-1)=0.4 \times \alpha_{LF}$. In
this paper we discuss LFs expressed in magnitudes.\\
Mass {\sl functions} are defined similarly.} (LFs, MFs).

For old GC systems in the local Universe, both LFs and MFs are of
log-normal (``Gaussian'') shape, with very similar parameters among a
wide variety of galaxies
(\citealt{1998gcs..book.....A,1991ARA&A..29..543H}; minor trends with
metal content can easily be accounted for, cf.
\citealt{1995AJ....110.1164A}), with a Gaussian peak (or ``turnover'')
magnitude at $M_V = -7.3$ mag and a Gaussian FWHM of 2.8 mag,
corresponding to a Gaussian $\sigma = 1.2$ mag. This universal
turnover of GC LFs is often used as ``secondary'' long-range
extragalactic distance indicator, and hence for determinations of the
Hubble constant and, as a consequence, of the expansion rate of the
Universe
(\citealt{1995ApJ...446....1S,1996AJ....112..954F,2000ApJ...533..125K}).

Young star clusters (YSCs, often referred to as ``open'' or ``populous''
clusters) in the local Universe seem to be of a different character
than the ubiquitous GCs. In the Milky Way, young clusters are mainly
sparse, low-mass objects ($\sim 10^3 - 10^4$ M$_\odot$), with low
concentration and lifetimes of order $10^8$ yr, that will soon
dissolve into the field star population. The Galactic open cluster
compilation of \citet{1984AJ.....89.1822V} suggests a LF of power
law-type with slope $\alpha \lta 0.5$ (but it might be steeper due to
incompleteness effects). The only local galaxies with a large number
of young clusters with masses $\gta 10^4$ M$_\odot$ (up to masses
$\lta 3 \times 10^5$ M$_\odot$) are the Magellanic Clouds. Work by
\citet{1985PASP...97..692E}, \citet{1997ApJ...480..235E} and
\citet{2003AJ....126.1836H} suggested a luminosity spectrum with a
power law of the form $N_{\rm YSC}(L) {\rm d} L \propto L^{\alpha}
{\rm d} L$, where $N_{\rm YSC}(L) {\rm d} L$ is the number of young
star clusters (YSCs) with luminosities between $L$ and $L + {\rm d}
L$, with $\alpha \approx -2$ (corresponding to a LF with $\alpha
\approx -1$). Recently, this power-law result was challenged by
\citet{2006MNRAS.366..295D}, who find significant deviations from a
power law for clusters in the Large Magellanic Cloud (LMC) of ages
younger than $\simeq$ 100 Myr and masses below $\simeq 3 \times 10^3$
M$_\odot$.

Rich systems of YSCs are routinely observed in starburst galaxies, in
particular in strong starbursts triggered by gas-rich galaxy
interactions and mergers (e.g.,
\citealt{1993AJ....106.1354W,1998AJ....116.2206S,1999AJ....118..752Z,
2001ApJ...561..727Z,2001AJ....121..768D,
2003NewA....8..155D,2003MNRAS.343.1285D}). Merger remnants with
post-starburst signatures also reveal SC systems with ages of up to
1--3 Gyr (e.g., \citealt{2003ApJ...583L..17D,2004ApJ...613L.121G}),
indicating that at least some fraction of the SCs formed during a
merger-induced burst survived much longer than most of the open
clusters in the Milky Way. The LFs of these YSC systems are usually
adopted to be of power law-type (see e.g. 
\citealt{2003MNRAS.343.1285D} for a recent compilation), but see
\citet{1998A&A...336...83F,1999A&A...342L..25F}, as well as 
\citet{2003ApJ...583L..17D} and \citet{2004ApJ...613L.121G} for
Gaussian-shaped LFs of young and  intermediate-age cluster systems,
respectively.

The lower maximum masses of the YSCs in the Milky Way as compared to
YSCs in the LMC and in starburst/merging galaxies could be (partially)
understood by the (purely statistical) size-of-sample effect, studied
e.g. by \citet{2003AJ....126.1836H}: In the latter galaxies the total
number of YSCs is larger than in the Milky Way, therefore the MFs are
sampled up to higher masses.

It has hitherto remained unclear whether the difference in shape
between the power-law LFs of young SCs (but see
\citealt{1998A&A...336...83F,1999A&A...342L..25F},
\citealt{2005A&A...433..447C}, \citealt{2003ApJ...583L..17D} and
\citealt{2004ApJ...613L.121G} for the Gaussian LFs in the Antennae
galaxies [all clusters], NGC 5253 [clusters  older than 10 Myr], M82B
[contains clusters at roughly 1 Gyr] and NGC 1316 [contains clusters
at roughly 3 Gyr], and the new results by
\citealt{2006MNRAS.366..295D} on the LMC SC system) and old GC systems
is caused by differences in the nature and formation of the two types
of clusters, or whether the power law of young systems is secularly
transformed into the Gaussian distribution of old GC systems by
selective destruction effects.

\subsection{Our case study: the Antennae system}

In this paper we study the YSC system of the Antennae galaxies (NGC
4038/39), the best-studied and nearest example of a major interaction
between two massive gas-rich spiral galaxies, at a distance of 19.2
Mpc \footnote{Throughout this paper we adopt a distance to the Antennae
galaxies of 19.2 Mpc ($m-M=31.4$ mag; see e.g.
\citealt{2002AJ....124.1418W,2003AJ....126.1276K,2004ApJ...605..725M}).
We point out that adopting the distance of 13.8 Mpc ($m-M=30.7$ mag)
suggested by \citet{2004AJ....127..660S} would result in an overall
shift of {\sl all} absolute magnitudes by 0.7 mag towards fainter
magnitudes, leaving the arguments presented in this paper
unaltered.}. In addition, the low inclination of both galaxies favours
the detection and study of their SC systems. In the Antennae galaxies,
both the individual SCs (see e.g.  \citealt{1995AJ....109..960W,
1998A&A...336...83F,1999A&A...342L..25F, 2001ApJ...561..727Z,
2002AJ....124.1418W, 2005ApJ...631L.133F}; for SCs in the extended
tidal tails see \citealt{2003AJ....126.1227K}) and the SC complexes
(see e.g.  \citealt{2005AJ....130.2104W,2006A&A...445..471B}) have
been studied extensively. In addition, the Antennae galaxies are a
testbed for a large number of studies related to star formation and
dynamical evolution in such a complex environment (see e.g.
\citealt{1988ApJ...331..699B,1998A&A...333L...1M,2001AJ....122.2969H,
2003ApJ...598..272F, 2004ApJ...605..725M, 2004ApJS..154..193W,
2005ApJ...619L..87H}).

The LF of the YSCs in the Antennae galaxies have been studied
extensively. However, so far no consensus about the shape of the LF
has been reached: see \citet{1999A&A...342L..25F} in favour for a
Gaussian shape, \citet{1999AJ....118.1551W} and
\citet{2005A&A...443...41M} for a broken-power-law shape (this could
be interpreted as intermediate case, with the shallower slope at the
faint end equivalent to the Gaussian approaching the turnover,
although \citealt{2006A&A...446L...9G} interpret this broken
power-law as a sign of a physical upper mass limit of star clusters)
and \citet{1995AJ....109..960W,1999ApJ...527L..81Z} in favour for a
power law shape.

Establishing whether or not the LFs of SC systems are the same or
different in quiescently star-forming galaxies compared to
merger-induced starbursts has far-reaching implications for our
understanding of the star formation and star cluster formation
processes themselves and their presumed universality, and possibly of
their environmental dependence. To assess whether or not there is a
clear turnover in the SC LFs (Section \ref{sec:turnover}), we
introduce statistical methods to model the sample selection procedure
(Section \ref{subsection:statmodel}), which includes parametric
modelling of the completeness function (our completeness studies for
the Antennae's SC system are summarised in Section
\ref{sec:completeness}) and, in a second step, a maximum likelihood
procedure to estimate the parameters of a power-law and a Gaussian
distribution, respectively, from the data as modified by this
completeness function. Applying a likelihood ratio test and
Monte-Carlo simulations to test its significance we find statistically
significant evidence for the presence of a Gaussian-like
turnover\footnote{i.e. at a confidence higher than 99.5 per cent (in
fact, {\sl none} of the 1000 simulated datasets shows a superiority of
the Gaussian over the power law as strong as in the real data)} in the
LF at absolute magnitudes between $-9.5$ and $-8.0$ mag in the $U, B$,
and $V$ bands, and at $\approx -6.5$ mag in the $I$ band (although
less significant, at a 68 per cent level). These results are presented
in detail in Section \ref{subsection:statresults}.

To investigate the robustness of these results, in Section
\ref{sec:subsets} we study subsets of the full sample, divided
according to various criteria. Although details change, the
conclusions are confirmed by these tests. The implications of our
results are discussed in Section \ref{sec:discussion}.

\section{Observations and data reduction}
\label{sec:obs}

We reanalyse the most homogeneous dataset of broad-band imaging
observations of the Antennae system available to date, obtained using
the {\sl Hubble Space Telescope (HST)}/Wide Field and Planetary
Camera-2 (WFPC2) as part of programme GO-5962 (PI B. Whitmore; see
Table \ref{tab:obs}). The images were retrieved via the on-the-fly
data reduction pipeline OPUS from the {\sl HST} data archive at the
Space Telescope European Coordinating Facility (ST-ECF).

\begin{table}
\caption{Observation log. All data obtained in Cycle 5 (in January
1996), using the {\sl HST}/WFPC2 camera.}
\begin{center}
\begin{tabular}{@{}*{3}{|c}{|}@{}}
\hline
Filter & Johnson equiv. passband & total exposure time (s)\\
\hline
F336W & $U$ & 4500 \\
F439W & $B$ & 4000\\
F555W & $V$ & 4400\\
F814W & $I$ & 2000\\
\hline
\end{tabular}
\label{tab:obs}
\end{center}
\end{table}

The image alignment and cosmic-ray rejection were done using standard
tasks in {\sc iraf}, including tasks from {\sc stsdas}\footnote{The
Image Reduction and Analysis Facility (IRAF) is distributed by the
(U.S.) National Optical Astronomy Observatories, which is operated by
the Association of Universities for Research in Astronomy, Inc., under
cooperative agreement with the (U.S.) National Science Foundation.
{\sc stsdas}, the Space Telescope Science Data Analysis System,
contains tasks complementary to the existing {\sc iraf} tasks.}.

The source identification was done using a version of {\sc DAOphot},
adapted to run under {\sc idl}. Since we do not {\sl a priori} know
whether or not the potentially young SCs we are seeking are
dynamically relaxed, we relaxed the roundness and sharpness criteria
slightly compared to the default values. We compared the number of
objects found using flux thresholds of $3 \sigma_{\rm bg}$ and $4
\sigma_{\rm bg}$, where $\sigma_{\rm bg}$ represents the r.m.s. of the
background flux, but found virtually no difference. We used the $> 4
\sigma_{\rm bg}$ sources, because we expected other
completeness-limiting aspects of the data analysis to have more severe
impacts (see \ref{sec:completeness}). Subsequently, we considered only
sources present in either all of the $UBV$, or in all of the $BVI$
images. This procedure prevents excluding either severely extincted
clusters (possibly lacking $U$-band imaging) or extremely young
objects (possibly lacking $I$-band data).

In total, we identified 7817 sources with fluxes above the threshold
value in either of our three adjacent passband sets. However, as
already shown by \citet{1999AJ....118.1551W} the majority of these
sources are contaminating bright stars in the Antennae galaxies
themselves. This is also clear from the estimated youth of the
starburst, with a strong component within the last few $\times 10^7$
yr, as is apparent from the presence of a large number of H{\sc ii}
regions (e.g. \citealt{2000AJ....120..670N}) and the observation of
H$\alpha$ emission (e.g. \citealt{1999AJ....118.1551W}, see their
Fig. 4). The number of bright supergiant-type stars, Luminous Blue
Variables, and bright O-type stars is difficult to estimate, but it is
expected to be large, with a fair number reaching absolute $V$-band
magnitudes of $\approx -9$ mag, and a few rare, Eta Carinae-type
objects reaching even $M_V \approx -10$ mag. However, the further
cluster selection criteria we will apply remove the vast majority of
these because of their properties, particularly the extendedness
criterion.

\subsection{Photometry and cluster sizes}
\label{sec:photom_sizes}

The crowding of the sources, their large number, partially their
spatial extent and the strong variations in the galactic background
contribution render photometry of the clusters extremely difficult.
While the large number of sources requires some automation in source
photometry and cluster candidate selection, this is severely hampered
by the other factors indicated above.

Therefore, we developed an automated ``cookbook'' to improve aperture
photometry by taking the sizes of the sources into account. For a
number of cluster light profiles and a wide range of intrinsic
cluster sizes we determined the ``observed'' size (as broadened by
the point-spread function [PSF] and the diffusion kernel of the
camera; the fitted profile used was a Gaussian, for simplicity and
stability reasons) as well as the size-dependent aperture corrections
(ACs) required to account for all of the cluster light (for a full
description see \citealt{2006A&A...451..375A}; hereafter
\citetalias{2006A&A...451..375A}). In the present paper we will apply
the algorithms presented and extensively tested in
the \citetalias{2006A&A...451..375A} to the Antennae data.

To each of the 7817 sources we fit a Gaussian light profile, as
justified and validated in the \citetalias{2006A&A...451..375A}, and obtain aperture photometry
using 3-pixel source apertures, and 5/8-pixel inner/outer radius sky
annuli. These sizes were chosen as a good compromise between being
small enough to avoid crowding and the potentially detrimental effects
caused by the highly variable background, and large enough to average
out the Poissonian noise from the source flux and aperture centring
effects.

 Since the majority of the clusters are younger than 25 Myr (see
\citealt{1999AJ....118.1551W} and Section \ref{sec:ages}), we do not expect 
them to have already
developed tidally-truncated King profiles \citep{1962AJ.....67..471K}.
We therefore assume the average profile representing the YSCs in the
LMC, namely an \citet{1987ApJ...323...54E} (EFF) profile, with a
power-law slope of $-3$, corresponding to an EFF15 profile in the {\sc
BAOlab} software environment\footnote{{\sc BAOlab} is a powerful image
analysis suite, described in \citet{1999A&AS..139..393L}. It is
especially powerful for cluster size measurements of marginally
resolved objects, and for the creation of artificial cluster images,
and thus for detailed completeness tests.}. If the measured sizes in
the different passbands differed, the luminosity-weighted mean of the
measurements was taken as representative size. Throughout this paper
we will use the FWHM of the EFF15 model as measure for the ``size'' of
the objects, unless otherwise stated.

Using the size information just obtained and the aperture photometry
of the clusters, we apply the recipes developed in the \citetalias{2006A&A...451..375A}, including
size-dependent ACs and background-oversubtraction corrections. The
total brightnesses were corrected for Galactic extinction
(\citealt{1998ApJ...500..525S}), converted from the {\sc stmag} system
of the {\sl HST} observations to the {\sc vegamag} system\footnote{The
additive offsets required are solely filter-dependent, as they result
only from the different calibration spectra. They can be retrieved
from the authors.}, and subsequently transformed to absolute
magnitudes.

For a source to be included in our final cluster sample it has to
fulfil the following criteria:

\begin{itemize}

\item It must have been found by the {\sc DAOphot}-like {\sc idl}
routine.

\item It must be detected at least in either all $UBV$ or all $BVI$
images, to filter out spurious detections and residual cosmic rays. In
addition, since we want to determine physical parameters for these
clusters using the {\sc AnalySED} algorithm
(\citealt{2004MNRAS.347..196A}) in a subsequent paper (Paper II), we
need multi-passband photometry, ideally for a large number of
passbands and a wide wavelength range.

\item The size determination must have converged. We need to correct
our photometry using the size-dependent ACs derived in
the \citetalias{2006A&A...451..375A}. Therefore, size information is
essential. However, the size determination becomes increasingly
difficult and unreliable for faint objects, hence also reducing the
sample's completeness (see Section \ref{sec:completeness}).

\item The converted (= {\sl intrinsic}) FWHM must be in the range from
  0.5 to 10 WF3 pixels (i.e., pixels on the wide-field-3 chip of the
  WFPC2 camera), corresponding to a FWHM of 4.6 to 93 pc (or a
  half-light radius $R_{1/2}$ of 5.2 to 105.1 pc, assuming an average
  young LMC ``EFF15''-type cluster profile) at the adopted distance of
  the Antennae. This range covers the range for which the
    formulae derived in the \citetalias{2006A&A...451..375A} can be
    applied confidently.  For clusters smaller than 0.5 WF3 pixel the
    correction equations determined in the
    \citetalias{2006A&A...451..375A} become inaccurate.  In addition,
    the lower size cut-off at 0.5 pixel (imposed to ensure that the
    source is extended, and hence a likely cluster) significantly
    reduces the sample contamination by bright supergiants, which
    would appear as point-like objects: the observed size would be on
    the order of the filter-dependent PSF size ($\sim 1.5-2$ pixels).
    The upper size limit was chosen as a compromise in view of the
    contamination by cluster complexes and the number of sources. To
    satisfy both constraints we will, where appropriate, distinguish
    between SC samples characterised by sizes in the range from
    $R_{1/2} \simeq 5$ to 25 pc (``small'' clusters) and our full
    sample. Note that cluster complexes in M51 and the Antennae have
    diameters of $\gta 100$ pc (see
    \citealt{2005A&A...443...79B,2006A&A...445..471B}).

\item The photometric error in each passband must be $\le 0.2$ mag,
because we need accurate cluster photometry to determine the physical
cluster properties within reasonable uncertainty ranges in Paper II.
\end{itemize}

The final full sample contains 752 clusters satisfying all of these
constraints, the small cluster sample contains 365 clusters.

\subsection{Completeness determination}
\label{sec:completeness}

In particular for a situation as complex as in the Antennae system,
completeness determinations are as difficult as cluster selection and
photometry.

The ultimate goal for the determination of the completeness function
of a cluster sample is the determination of a local completeness
fraction for each cluster, by taking into account all effects of the
local environment of each cluster and all relevant selection effects.

In reality, at least the detailed spatial dependence of the
completeness functions can hardly ever be determined accurately for
each cluster. Here, we determine the completeness functions for two
different cluster sizes (for intrinsic sizes of 1 and 2 pixels,
chosen, respectively, to represent the size bin with the largest
number of clusters, and roughly approximating the median cluster size)
in two distinct regions of the galaxies, distinguished by their
average source density. The location of the regions are shown in Fig.
\ref{fig:regions}. We labelled the regions {\sc disc} region (on the
left-hand side, in the disc of NGC 4038) and {\sc overlap} region (on
the right-hand side, in the region where the discs are overlapping,
and the presence of large amounts of dust is seen in white).

\begin{figure}%[h]
 \begin{center}
 \includegraphics[angle=0,width=0.9\columnwidth]{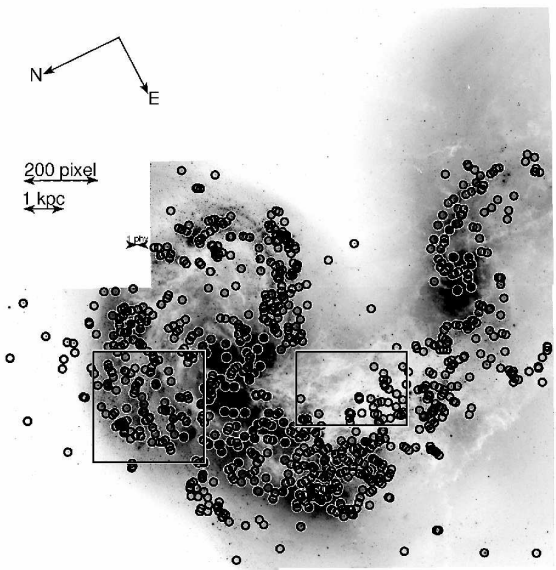}
 \end{center} 
 \caption{The Antennae galaxies, with the two regions for which the
 completeness functions were determined indicated: on the left-hand
 side the {\sc disc} region, on the right-hand side the {\sc overlap}
 region. The clusters are circled. Orientation and length scales
 indicated in the image.}
 \label{fig:regions} 
\end{figure}

For both regions, we performed a number of tests to estimate the
completeness functions. For the {\sc disc} region we will present the
different methods and compare their results (see Figs.
\ref{fig:star_comp} and \ref{fig:cluster2_comp}). For the most
sophisticated method (taking into account filter cross-correlation, 
photometric accuracy and the requirement for the cluster size determination
to converge) we will show the results for the {\sc overlap}
region as well (see Fig. \ref{fig:cluster2_comp}). The results are
summarised in Table \ref{tab:completeness}.

The simplest way to estimate completeness functions is by distributing
Gaussian-shaped artificial sources (``stars'') with a range of
brightnesses onto the image, e.g. by using the {\sc iraf} routine {\sc
mkobj} with radii corresponding to the PSF sizes, and then to
determine the fraction of sources being recovered by the source
finding algorithm.  This can be done for each passband separately (see
Fig.  \ref{fig:star_comp}, upper panel, thin solid lines with
symbols).  However, for source identification we require that a
detection must be positive in at least 3 adjacent passbands out of the
4 available. For this cross-correlation, a spectral energy
distribution (SED) must be assumed. We investigated the cases of a
flat SED (all colours are equal to 0, by definition), an SED
representative of a young solar-metallicity cluster (age = 12
Myr, $Z=0.02$ = Z$_\odot$), and one for an intermediate-age
subsolar-metallicity cluster (age = 100 Myr, $Z=0.008$ = 0.4
Z$_\odot$); see the thin long-dashed lines in the upper panel of
Fig. \ref{fig:star_comp} for the results.

In addition, we require the clusters to have good photometry, with
photometric errors smaller than 0.2 mag in each passband. This
requirement further decreases the completeness fractions. The results
are shown in the upper panel of Fig. \ref{fig:star_comp} by the thick
dot-dashed lines.

\begin{figure}%[h]
 \begin{center}
 \includegraphics[angle=270,width=0.95\columnwidth]{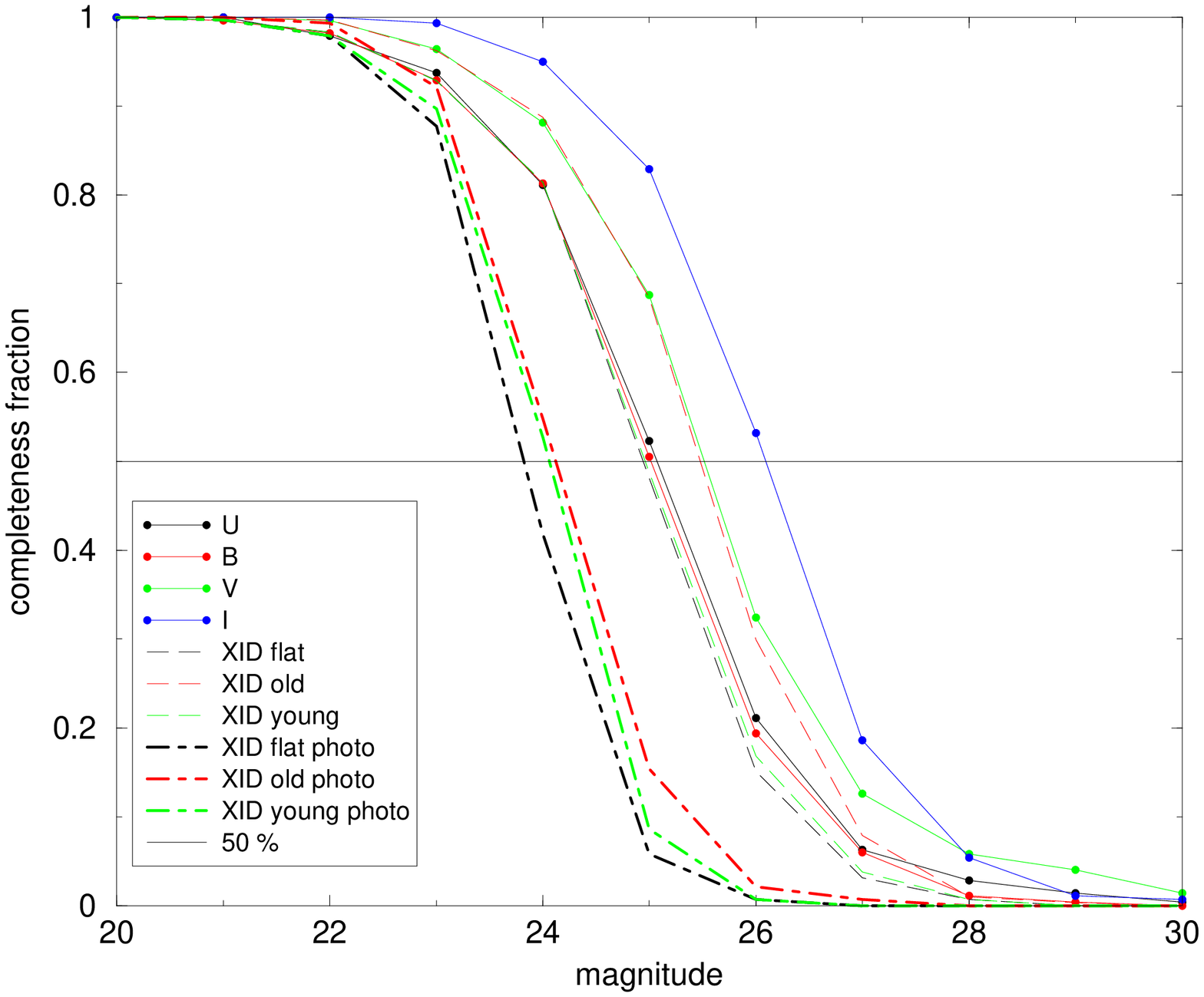}
 \includegraphics[angle=270,width=0.95\columnwidth]{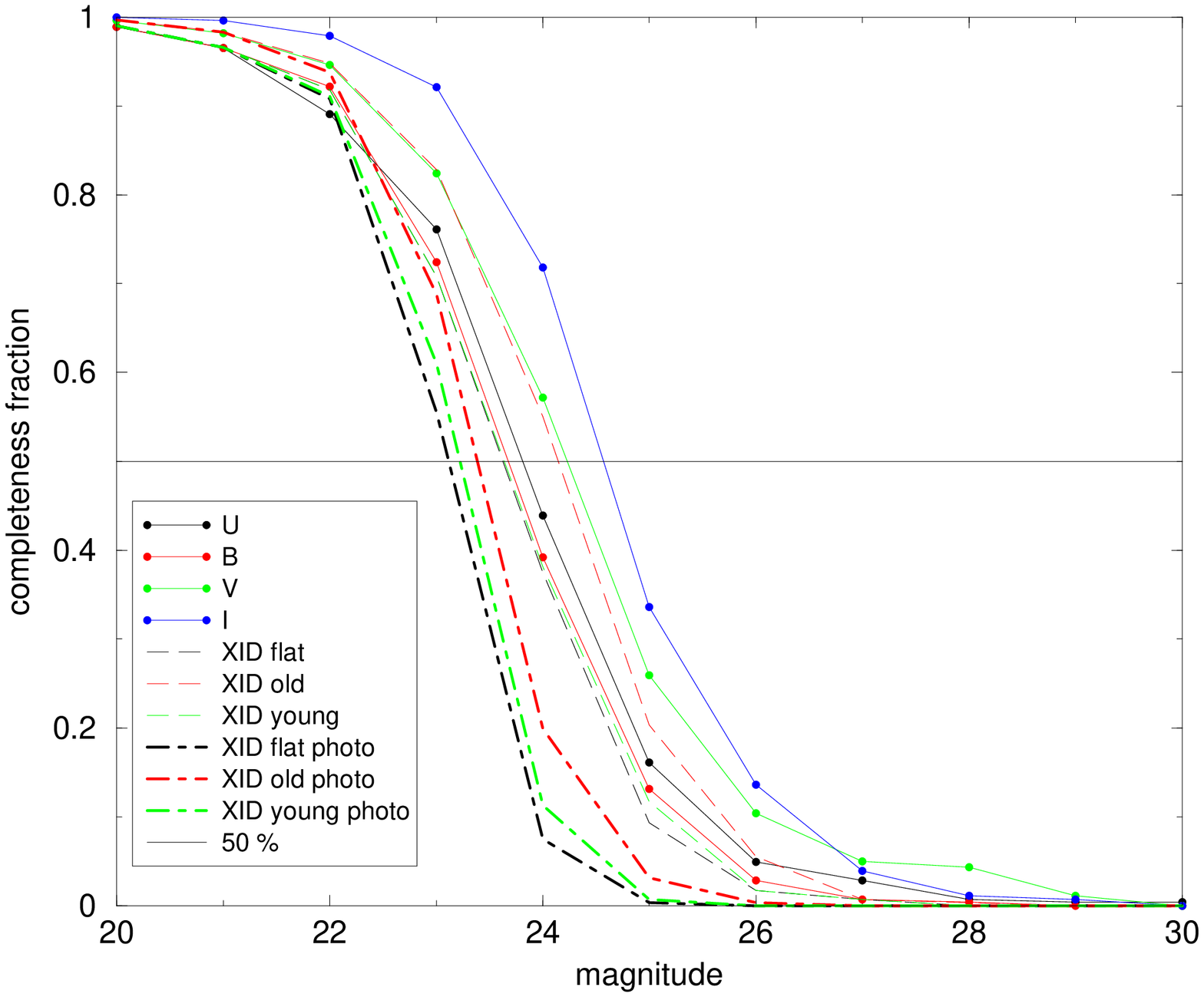}
 \end{center} 
 \vspace{0.6cm}
 \caption{Upper panel: Completeness fractions based on artificial {\sl
 stars}. Bottom panel: Completeness fractions based on artificial {\sl
 clusters} with FWHM = 1 pixel. Thin solid lines with bullet points: Completeness
 functions for the individual filters. The horizontal axis is the
 magnitude in the respective passband. Thin dashed lines: Completeness
 functions for cross-correlating the filters. The horizontal axis is
 the magnitude in the $V$ band. Thick dot-dashed lines: Completeness
 functions for cross-correlating the filters and requiring good
 photometry (uncertainties $\le 0.2$ mag). The horizontal axis is the
 magnitude in the $V$ band. XID = cross-correlation. Both panels relate to the 
 DISC region.}
 \label{fig:star_comp} 
\end{figure}

However, neither the Gaussian shape nor the small size of the
artificial stars are realistic representations of a YSC. We therefore
built more realistic cluster models (with intrinsic FWHM = 1 and 2
pixels, as explained above), using {\sc BAOlab}, taking both the
appropriate {\sl HST}/WFPC2 PSFs (based on the {\sc Tiny Tim} software
package; see \citealt{krist97}) and the {\sl HST}/WFPC2 diffusion
kernel into account. We then performed the detection,
cross-correlation and photometry tests once again. The results are
shown in the bottom panel of Fig. \ref{fig:star_comp}.

For the real clusters an additional constraint imposed is the size
determination. Hence for each cluster the size determination has to
converge. Since the size fitting will not converge easily for faint
clusters, this further reduces the completeness fraction of the source
detections. In addition, all methods are sensitive to the cluster size.
These results are given in Fig. \ref{fig:cluster2_comp}, which also
provides the comparison between the completeness determined in the {\sc
disc} region with that for the {\sc overlap} region.

\begin{figure}%[h]
 \begin{center}
 \includegraphics[angle=270,width=0.95\columnwidth]{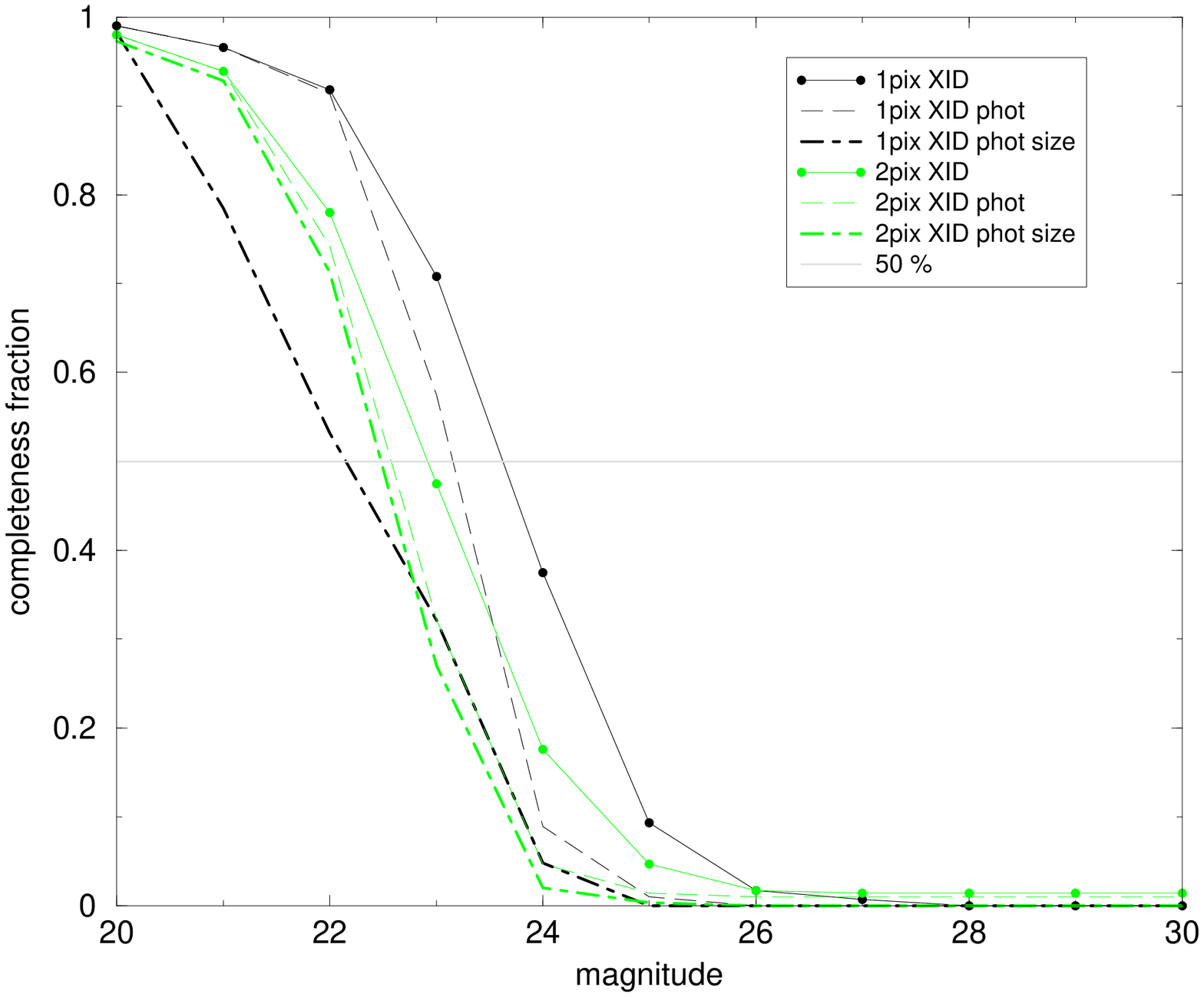}
 \includegraphics[angle=270,width=0.95\columnwidth]{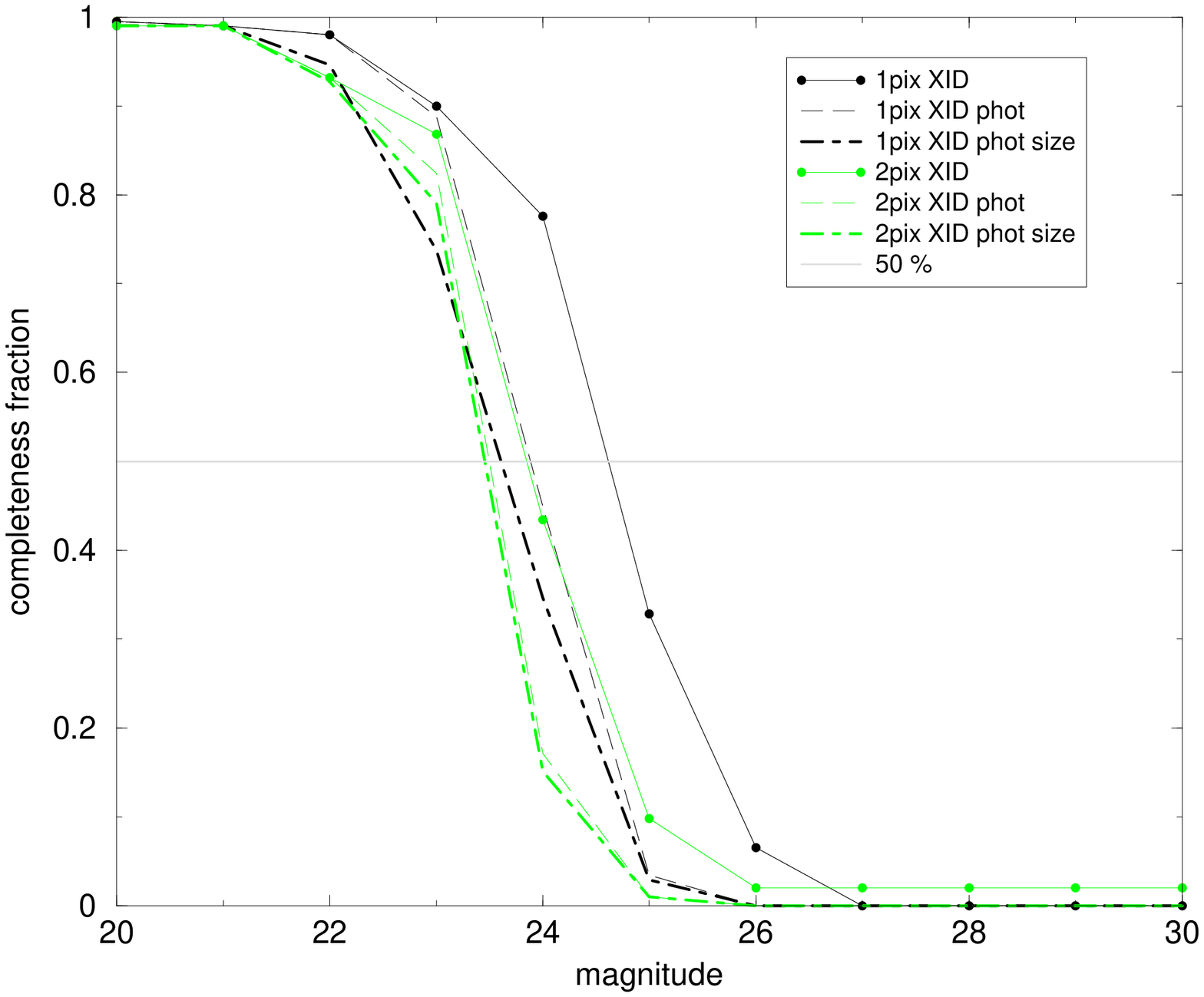}
 \end{center} 
 \vspace{0.6cm}
 \caption{Completeness fractions based on artificial clusters, taking
 the cluster sizes and size determination effects into account.  Black
 lines correspond to clusters with intrinsic FWHM = 1 pixel, light
 grey lines (in colour print: green lines) to clusters with intrinsic
 FWHM = 2 pixels. Solid thin lines with bullet points: only
 cross-correlation (here only the cross-correlations using the flat
 SED are shown, for reasons of clarity).  Long-dashed thin lines:
 cross-correlation including photometric accuracy determination; thick
 dot-dashed lines: full completeness fractions including size
 determination. Upper panel: {\sc disc} region, bottom panel: {\sc
 overlap} region. XID = cross-correlation.}
 \label{fig:cluster2_comp} 
\end{figure}

\begin{table}
\caption{Comparison of 50 per cent completeness limits for artificial
sources of different types (stars versus clusters; 2 cluster sizes),
different regions within the Antennae system, and different methods of
completeness determination. XID = cross-correlation.}
\begin{center}
\begin{tabular}{@{}*{6}{|l}{|}@{}}
\hline
model & star & cluster & cluster & cluster & cluster \\
 & {\sc disc} & {\sc disc} & {\sc disc} & {\sc overlap} & {\sc overlap}\\
 & & 1 pix & 2 pix & 1 pix & 2 pix \\
\hline
\multicolumn{6}{|c|}{\textbf{50 per cent completeness limits in individual passbands}} \\
\hline
 $U$ only & 25.1 & 23.8 & 23.2 & 24.6 & 24.0\\
 $B$ only & 25.0 & 23.7 & 23.1 & 24.8 & 24.1\\
 $V$ only & 25.5 & 24.2 & 23.6 & 25.3 & 24.5\\
 $I$ only & 26.1 & 24.6 & 24.1 & 25.2 & 24.6\\
\hline
\multicolumn{6}{|c|}{\textbf{50 per cent completeness limits of cross-correlated data}} \\
\hline
 XID only & 24.9 & 23.6 & 22.9 & 24.6 & 23.8\\
 phot & 23.8 & 23.2 & 22.6 & 23.9 & 23.5\\
 phot + size & $-$ & 22.2 & 22.5 & 23.6 & 23.5\\
\hline
\end{tabular}
\label{tab:completeness}
\end{center}
\end{table}

\section{Luminosity functions}
\label{sec:luminosity}

Fig. \ref{fig:lf} shows the cluster LFs, i.e. the number of clusters
per magnitude bin, for the four passbands. While the LFs appear to
exhibit a turnover, the decrease at the faint end is (at least
partially) caused by the observational incompleteness. To separate the
effects of (in)completeness from a possible intrinsic decrease in the
number of faint clusters (i.e. an intrinsic turnover), we designed the
statistical tools described in Section \ref{sec:turnover}.

\begin{figure}%[h]
 \vspace{0.5cm}
 \begin{center}
 \includegraphics[angle=0,width=0.95\columnwidth]{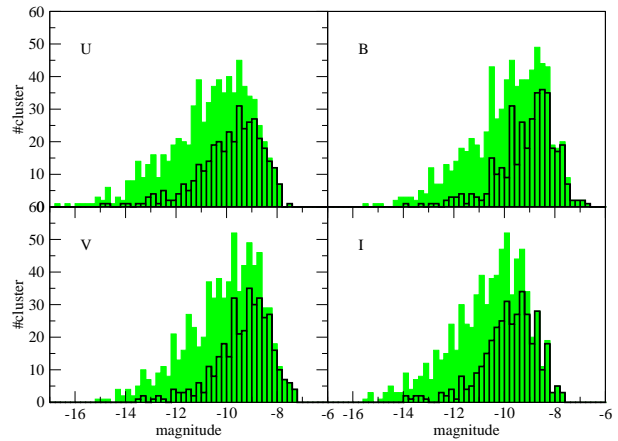}
 \end{center} 
 \caption{LFs of the final cluster sample in the four passbands, not
 corrected for the effects of sample (in)completeness. The shaded
 histogram shows the full cluster sample, the open histogram
 represents the small cluster sample.}
 \label{fig:lf} 
\end{figure}

\subsection{Comparison with data in the literature}

The pioneering work of Whitmore and collaborators has resulted in the
largest homogeneous dataset of star cluster photometry for the Antennae
system. For a direct comparison Brad Whitmore kindly provided us with
the photometry of their full (unpublished) source sample.

\begin{figure*}%[hb!]
 \vspace{0.5cm}
 \begin{center}
 \hspace{-5.0cm}
 \includegraphics[angle=0,width=\textwidth]{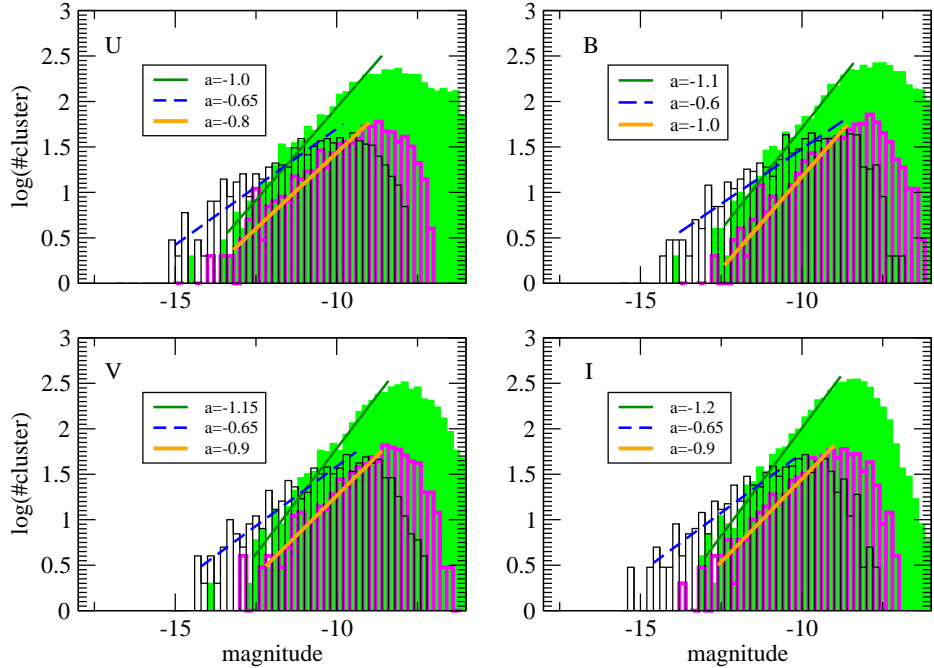}
 \end{center} 
 \vspace{-0.4cm}
 \caption{LFs of the final cluster sample in the four passbands,
 compared to the corresponding data from
 \citet{1999AJ....118.1551W}. The slopes, $a$, of the linear fits to
 the logarithms of the LFs are indicated in the legends. This slope 
 is equivalent to the LF power-law slope $\alpha_{\rm LF}$.}
 \label{fig:lfcomp1} 
\end{figure*}

As can be seen from Fig. \ref{fig:lfcomp1}, our LFs are substantially
flatter than those of the Whitmore sample. This can be partially
attributed to the effect of the ACs: bright clusters in our sample
tend to be larger (i.e. have larger ACs), which flattens the LF.
Overall, our sample without the ACs is in better agreement with the
Whitmore sample (despite the vastly different sample sizes). In
addition, the different selection criteria probably also play an
important role in the slope differences, as our selection criteria
largely exclude the stellar contamination discussed in
\citet{1999AJ....118.1551W}, which is expected to dominate the faint
end of the LF and thus will lead to a steepening of the LF.

\section{Turnover determination}
\label{sec:turnover}

Particularly important for our study of the LFs is the potential
existence and the location of a turnover in our dataset to establish
the link between our YSCs and the old GCs. Unfortunately, the
completeness function drops off already at significantly brighter
magnitudes. This results in an apparent turnover of the {\sl observed}
cluster luminosity distribution for any reasonable {\sl intrinsic}
luminosity distribution (which we call the ILF), like a power-law or
Gaussian distribution. Therefore, the process that leads to the
observations, has to be modelled carefully before any conclusion can
be drawn with respect to the existence of a possible turnover.

In Section \ref{subsection:statmodel} we present a detailed
statistical model for the observations and develop an appropriate
statistical procedure to test for the presence of a turnover. In
section \ref{subsection:statresults} we apply these methods to the
Antennae YSC data.

\subsection{Modelling the observations}
\label{subsection:statmodel}

The observed luminosity distribution is different from the ''true'' 
ILF by virtue of two different stochastic mechanisms. First, the
observations are affected by measurement errors. Even more important
though is a so-called ``missing data'' mechanism described, in
essence, by the completeness function. Here, the probability of a
cluster to be included in the observations $X_i$ ($i=1,\ldots,N$)
depends on its brightness. The fainter the cluster, the less probable
it is to be observed. Moreover, if the cluster is fainter than $M_V
\approx -5$ to $-6$ mag, the completeness function approaches zero,
and it is not possible to analyse this part of the LF at all. For a
general introduction to statistical methods with missing data see,
e.g., \citet{little02}, for a biometric example see \citet{patil78}.

We now introduce the statistical model in more detail. Call
$f_{\vartheta}$ the ({\sl intrinsic}) LF, i.e. the (probability)
density of the brightness of a cluster, and assume that the intrinsic
luminosities of the Antennae clusters are realisations, $m_i$, of
independent random variables, $M_i$, distributed according to
$f_{\vartheta}$. Here, upper-case letters indicate random variables, 
and lower-case letters their realisations (i.e. the observed value). In 
the literature, essentially only log-normal (``Gaussian'') and power-law
models are under consideration for $f_{\vartheta}$, where $\vartheta$
are the parameters of the model, to describe the LFs of star clusters.

Please note: We are going to fit both the Gaussian and the
power-law models to the distribution of cluster magnitudes. Although
in the literature the power-law models are generally fitted to the
luminosity distributions, our method is a conservative approach, as
it assigns less weight to the  (potentially less well-defined) faint
end. Hence, if we can show that a power law model for the cluster
magnitudes predicts too many faint clusters, this would, a fortiori,
hold for a power law in cluster luminosities.

Even if the completeness function would be unity for all magnitudes,
the measurements would not be equal to the $m_i$, but scatter around
these true values due to measurement errors. We call $e(\cdot|M)$
the probability density of the measurement error of a cluster with
intrinsic brightness $M$. More formally, $e(\cdot|M)$ is the {\sl
conditional density} of the measurement errors given that the
intrinsic cluster brightness is $M$, i.e.  the probability density
that the measurement error has a certain value if we already know that
the intrinsic brightness is $M$. Then, the {\sl observable} brightness 
is given by the random variable $Y_i = M_i + E_i$, where the $E_i$ are
distributed independently of each other, as $e(\cdot|M_i)$. 
We call $Y_i$ the {\sl observable} brightness, as opposed to the {\sl
observations} $X_i$, since not all of the clusters (with {\sl observable}
brightness $Y_i$) will
actually be included in an observed sample due to the (in)completeness
effects.

The error density $e(\cdot|M)$ can, for example, be a Gaussian density
with fixed variance, i.e. independent of $M$, or a Gaussian with a
variance which depends on $M$. Since the observed values basically
result from photon count data, a Poisson distribution with 
expected value computed from $M_i$ is another straightforward
possibility. We use the latter choice. However, computations where we
assumed Gaussian noise yield very similar results. This is because the
measurement errors are small, and in particular modify the observed
cluster distribution much less than the missing data effect described
next.

The probability of a cluster with observable brightness $Y$ to be
included in the observations is given by the completeness function
$c(Y)$. We model this by a random variable $Z$ which is Bernoulli 
distributed with parameter $c(Y)$, 
i.e. $Z$ is $1$ with probability $c(Y)$, and $0$ otherwise.

Recapitulating, the observable brightness is modelled as pairs
$(Y_i,Z_i),\, i=1,\ldots,N$ of a random variable $Y = M + E$ with 
probability densities $f_{\vartheta}$ of $M$, $e(\cdot|M)$ of
$E$, and $g_{\vartheta}$ of $Y$, and a second random variable $Z$,
which is Bernoulli distributed with parameter $c(Y)$. 
The observations consist of those pairs $(Y_i,Z_i)$
where $Z_i=1$. This implies that the probability density $h$ of an
observation $X_i$ is the joint density $p_{\vartheta}$ of the pair
$(Y,Z)$ conditioned on the event $Z=1$: \[ h(x) = p_{\vartheta}(x|z=1)
= \frac{p_{\vartheta}(x,z=1)}{P_{\vartheta}(Z=1)}, \] where
$P_{\vartheta}(Z=1)$ is the (marginal) probability that $Z=1$.

In the remainder of this section we describe our estimator for the
best-fit values $\hat\vartheta^{\rm mod}$ of the parameter(s)
$\vartheta$ for the Gaussian and power-law models, respectively,
i.e. ``mod'' reads either as ``Gaussian'' or ``power law''. Our
approach is based on maximum likelihood estimation, which amounts to
maximising the probability of the observed data $(Y_i,Z_i=1),\,
i=1,\ldots,n$, given the model ``model'' with parameter(s)
$\vartheta$, i.e.

\[\hat\vartheta^{\rm mod} = {\rm argmax}_{\vartheta} {\cal
L}_{\vartheta}(X_1,\ldots,X_N), \] 

\noindent
where 

\[{\cal L}_{\vartheta}(X_1,\ldots,X_N) \equiv \prod\limits_{i=1}^n p^{\rm
mod}_{\vartheta}(X_i|z=1), \] 

\noindent
and $p^{\rm mod}_{\vartheta}(Y = X_i|z=1)$ is the (conditional)
probability of a cluster to have observable brightness $X_i$ given
$Z=1$, i.e. to be actually contained in the observations. To compute
this conditional density we first determine the marginal probability
that $Z=1$ as

\begin{eqnarray} P^{\rm mod}_{\vartheta}(Z=1) = 
\int\limits_{-\infty}^{\infty} p^{\rm mod}_{\vartheta}(z=1|y)
g^{\rm mod}_{\vartheta}(y) {\rm d}y &&\nonumber\\ = 
\int\limits_{-\infty}^{\infty} c(y)
\left[\int\limits_{-\infty}^{\infty} p^{\rm mod}_{\vartheta}(y|m)
f_{\vartheta}(m) {\rm d}m\right]{\rm d}y. & & \nonumber \end{eqnarray} 

\noindent 
Here we use that $p^{\rm mod}_{\vartheta}(z=1|Y) = c(Y)$ (i.e. does not
depend on the specific model at all) and

\[g^{\rm mod}_{\vartheta}(y)=\int\limits_{-\infty}^{\infty} 
e(y-m|m) f^{\rm mod}_{\vartheta}(m) {\rm d}m.\] 

\noindent
Here, $e(y-m|m)$ is the probability that $Y=y$ given $M=m$. 
Moreover,

\begin{eqnarray} p^{\rm mod}_{\vartheta}(x,z=1) = p^{\rm
mod}_{\vartheta}(z=1|x) g^{\rm mod}_{\vartheta}(x) & & \nonumber\\
= c(x) \int\limits_{-\infty}^{\infty} e(x-m|m)
f^{\rm mod}_{\vartheta}(m) {\rm d}m. & & \nonumber
\end{eqnarray} 

\noindent
Therefore, the likelihood function of the observations $X_1,\ldots,X_N$ 
is

\[
{\cal L}^{\rm mod}_{\vartheta}(x_1,\ldots,x_N) = \prod\limits_{i=1}^n \frac{c(x_i) 
\int\limits_{-\infty}^{\infty} e(x_i\!-\!m|m) f^{\rm mod}_{\vartheta}(m) {\rm d}m}
{P^{\rm mod}_{\vartheta}(Z=1)^n}.
\]

Maximisation of ${\cal L}^{\rm mod}_{\vartheta}(X_1,\ldots,X_N)$ yields the
estimators (``best-fit parameter(s)'') $\hat\vartheta^{\rm mod}$.
Moreover, we assign the likelihood to the respective model under
consideration as a measure of its ``probability'' given the data.
This can be used to select the best model, particularly by comparing
the ratio of the likelihoods of the best-fit Gaussian and power-law
model in a likelihood ratio test. Versions of this kind of model
selection are well established for a broad range of statistical
applications (see e.g. \citealt{lehmann94}).

\subsection{Statistical significance of a turnover in the ILF}
\label{subsection:statresults}

We now discuss the results of our application of this method to the
Antennae clusters. For simplicity, we parametrise the completeness
functions as Fermi functions,

$$
(1+\exp(a(x-b)))^{-1}\qquad a,b\in \mathbb{R}.
$$

\noindent
We emphasise that the general results do not change if we use, e.g.,
interpolation methods to describe the completeness function.

For the $V$ band, the completeness data as determined in Section
\ref{sec:completeness} (crosses) and the fitted parametrisation
(solid line) are shown in Fig.~\ref{fig:compl_fit}. Three completeness
fractions were determined per magnitude, corresponding to the three
different cluster SEDs considered, giving some uncertainty estimate.
The fit was performed as $L_1$ -- Fit (i.e. by minimising the absolute
distance between model function and simulated completeness fractions).

\begin{figure*} 
 \begin{tabular}{lcc}
 \parbox{3cm}{}& {\sc overlap} Region & {\sc disc} Region\\
 \parbox{3cm}{\raisebox{-1cm}{1 pixel clusters}}&
 \parbox{5cm}{\includegraphics{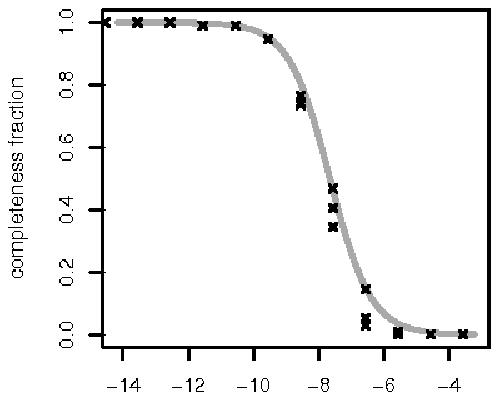}}&
 \parbox{5cm}{\includegraphics{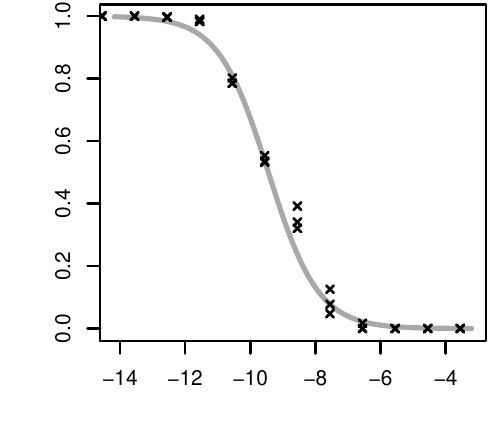}}
 \\
 \parbox{3cm}{\raisebox{-0.8cm}{2 pixel clusters}}&
 \parbox{5cm}{\includegraphics{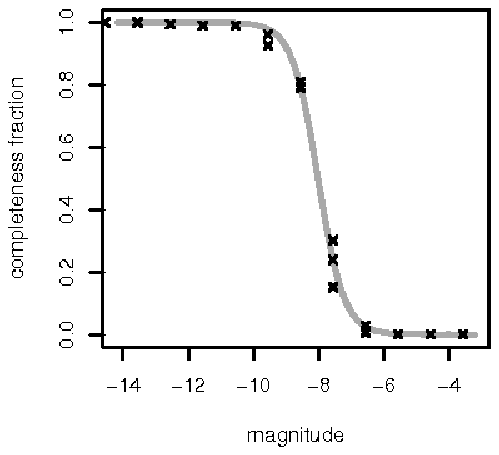}}&
 \parbox{5cm}{\includegraphics{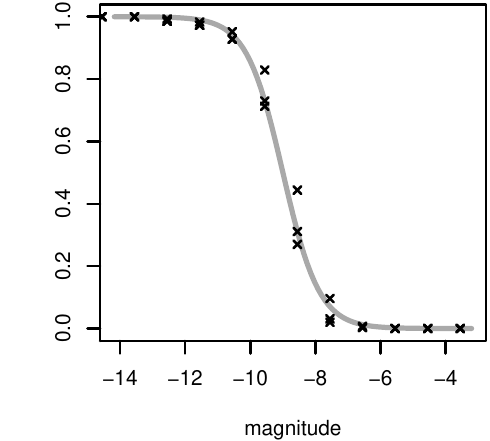}}
 \\
 \end{tabular}
 \caption{\label{fig:compl_fit} The completeness curve fits
 determined for the $V$ band. Crosses: Data points from completeness tests.
 Gray solid line: Fermi function fit to the data points. 
 Note that for different bands only the
 location on the magnitude axis varies but not the shape of the
 curve.}
\end{figure*}

Fig. \ref{fig:Model_fits} shows the best fits for the Gaussian and the
power-law models for the different passbands. A visual inspection of
Fig.~\ref{fig:Model_fits} already indicates that the intrinsic
Gaussian models fit better, particularly with respect to the faint
tail of the LF. The intrinsic power-law models fit quite well in all
bands for bright magnitudes but assign too much weight (i.e. predict
too many clusters) to the faint magnitudes. One might suspect that
this is due to the completeness curve assigning too much probability
to that magnitude range. However, an inspection of Fig.
\ref{fig:compl_fit} shows that for $-7 \le M_V \le -6$ mag, where the
power law predicts too many clusters, the completeness curves estimate
the generated completeness fractions without significant bias. This
observation can also be made for the other bands. Moreover, fitting a
different completeness curve (e.g., a linear interpolation of the
mean/median of the completeness fractions at each magnitude) leads to
the same effects. One has to keep in mind that the completeness
fraction plays a crucial role when fitting a model in this context.

\begin{figure*} 
 \begin{tabular}{lcc}
 \parbox{6.5cm}{\hspace{3.8cm}Gaussian} & Power Law
 \\
 \parbox{1.5cm}{$U$ band}
 \parbox{5cm}{\includegraphics{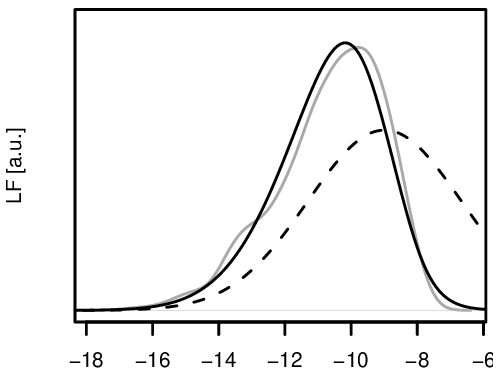}}&
 \parbox{5cm}{\includegraphics{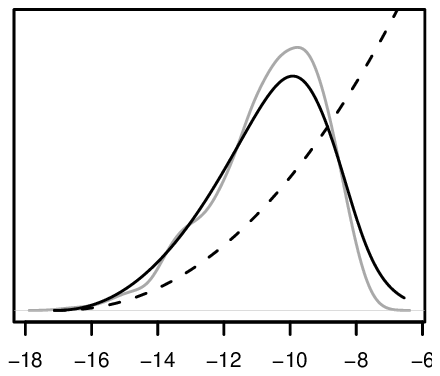}}
 \\
 \parbox{1.5cm}{$B$ band}
 \parbox{5cm}{\includegraphics{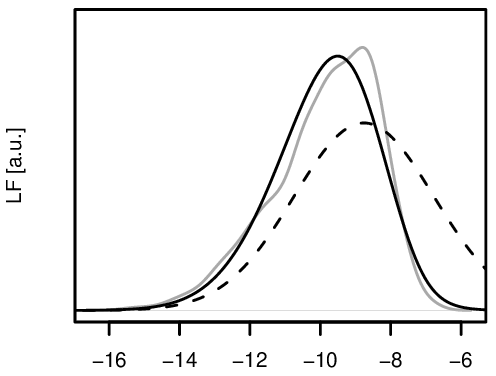}}&
 \parbox{5cm}{\includegraphics{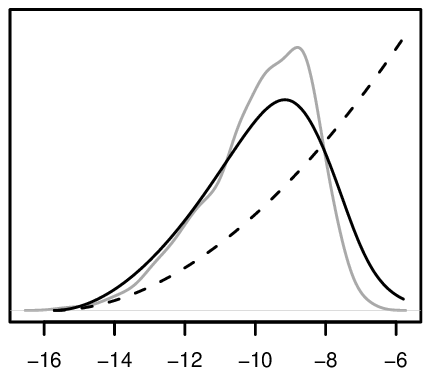}}
 \\
 \parbox{1.5cm}{$V$ band}
 \parbox{5cm}{\includegraphics{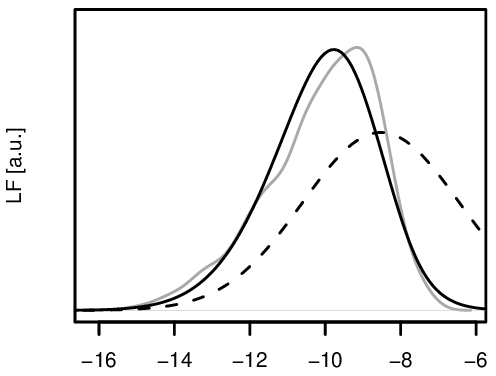}}&
 \parbox{5cm}{\includegraphics{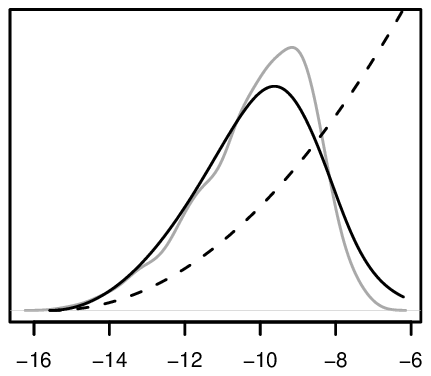}}
 \\ 
 \parbox{1.5cm}{$I$ band}
 \parbox{5cm}{\includegraphics{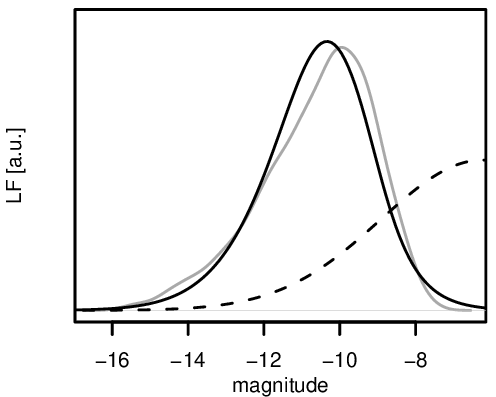}}&
 \parbox{5cm}{\includegraphics{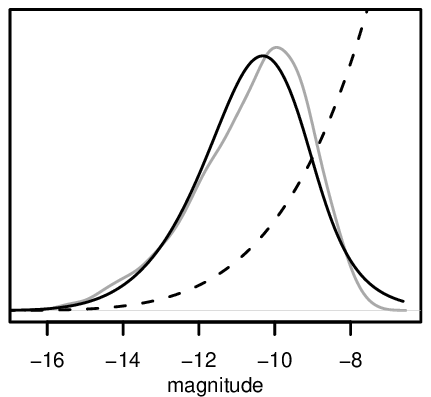}}
 \end{tabular}

 \caption{\label{fig:Model_fits} Fits of the Gaussian and the
 power-law model for the different passbands. The grey line is a
 kernel density estimate of the observed data and the black line shows
 the Maximum Likelihood fit multiplied by the weighted sum of the
 completeness curves (with weights chosen according to the relative
 frequencies of observations for the different regions and cluster
 sizes). The dashed line corresponds to the estimated LF without
 considering the completeness. The vertical scaling is in arbitrary
 units (a.u.), but scaled such that the modelled LF (i.e., the ILF convolved
 with the distribution of observational errors, multiplied with the
 completeness function) gives the same number of clusters as
 observed.}

\end{figure*}

More robust conclusions can be drawn using the model likelihoods. To
test the (null) hypothesis ``true model is power law with the
estimated best fit parameters'' against the alternative of a Gaussian
density we use a Likelihood Ratio test (cf. \citealt{lehmann94}).  We
first simulate the distribution of the ratio of the likelihoods of a
Gaussian and a power-law model under the hypothesis that the power-law
model holds true. This means that we randomly draw a number of
observations equal to the number of data point from the fitted
power-law model (using the fitted completeness curves) and then again
fit a power law and a Gaussian model and calculate the ratio of the
corresponding likelihoods. This procedure is repeated 1000 times and
the empirical quantiles of the resulting distribution are compared to
the observed likelihood ratio of the data. Table~\ref{tab:sim results}
shows that the probability to erroneously reject the null hypothesis
is at a level of less than 0.5 per cent for the $U$, $B$ and $V$ band,
i.e. the result is ``strongly significant'' (in fact, for most cases
considered {\sl none} out of $1000$ simulations, where the  ILF is
{\sl assumed} be a power-law, shows a superiority of the Gaussian over
the power-law as strong as in the real data). Note, that we {\sl
estimate} the probability of erroneously rejecting the null hypothesis
from the rate of rejecting the power law in favour of a Gaussian in
simulations with artificial data generated from the power law model.
Here, the result of zero rejects out of $1000$ simulations implies an
{\sl estimate} of the {\sl true rate of rejection} in the simulations
of $p\leq 0.5\%$,  with an error probability of this claim to be wrong
of $\lta 1\%$, based on the characteristics of Bernoulli-distributed
random variables.
For the $I$ band, the
observed value is greater than 74 per cent of the simulated values.
This means that the corresponding probability of erroneous rejection
(i.e.  the corresponding $p$-value) is 0.322 (i.e. 32.2 per cent) and
the null hypothesis cannot be rejected at any reasonable level.

In Table \ref{tab:sim_results_bestfit} we present the parameters 
and the uncertainty ranges from bootstrapping for the best-fitting
Gaussian models.

\begin{table}
 \caption{Results of the Likelihood Ratio tests. For clarity the
 difference of the negative log-likelihood values is given instead
 of the likelihood ratios. Hence, negative values indicate that the
 null hypothesis (i.e. power law) is superior. The last column gives
 the probability of error for discarding an intrinsic power law
 (hence small values represent the inferiority of the power law).}

 \begin{tabular}{|l| r r r r r r|}
 \hline
 & \multicolumn{3}{|c|}{\textbf{Quantiles}} & & & \\
Filter & 50\% & 95\% & 99\% & 100\% & {\bf Obs.} & Prob.\\
 \hline
$U$ & -4.5 & 0.1 & 2.5 & 5.0 & {\bf 15.5} & 0.000 \\
$B$ & -6.0 & -0.9 & 1.3 & 6.2 & {\bf 32.5} & 0.000 \\
$V$ & -5.4 & -0.05 & 2.5 & 9.4 & {\bf 15.3} & 0.000 \\
$I$ & -1.2 & 1.8 & 3.2 & 5.5 & {\bf -0.4} & 0.322 \\ 
 \hline
 \multicolumn{7}{|c|}{\textbf{FWHM=0.5-2.36 pixel $\rightarrow$ $r_{1/2} \simeq$ 5-25 pc}}\\
 \hline
$U$ & -0.9 & 2.6 & 5.0 & 6.8 & {\bf 18.5} & 0.000 \\
$B$ & -1.3 & 3.0 & 4.6 & 8.9 & {\bf 35.5} & 0.000 \\
$V$ & -0.5 & 4.0 & 5.9 & 8.9 & {\bf 30.8} & 0.000 \\
$I$ & -0.1 & 3.4 & 5.2 & 7.7 & {\bf 16.8} & 0.000 \\ 
 \hline
 \end{tabular}
 \label{tab:sim results}
\end{table}

\begin{table}
 \caption{Best-fitting parameters for the Gaussian model.
 ($\mu$ = mean)}
 \begin{tabular}{|l| r c r |}
 \hline
 & \multicolumn{2}{|c|}{\textbf{Gaussian}} & \\
Filter & $\mu$ & 95 percentile range $\mu$ & $\sigma$ \\
 \hline
 $U$ & -9.0 & $[$-9.4:-8.5$]$ & 2.3 \\
 $B$ & -8.8 & $[$-9.0:-8.5$]$ & 2.0 \\
 $V$ & -8.5 & $[$-8.8:-8.1$]$ & 2.1 \\
 $I$ & -6.2 & $[$-7.3:-5.0$]$ & 2.7 \\ 
 \hline
 \multicolumn{4}{|c|}{\textbf{FWHM=0.5-2.36 pixel $\rightarrow$ $r_{1/2} \simeq$ 5-25 pc}}\\
 \hline
 $U$ & -8.6 & $[$-8.9:-8.2$]$ & 1.7 \\
 $B$ & -8.3 & $[$-8.5:-8.0$]$ & 1.4 \\
 $V$ & -8.4 & $[$-8.6:-8.1$]$ & 1.4 \\
 $I$ & -8.0 & $[$-8.4:-7.5$]$ & 1.5 \\ 
 \hline
 \end{tabular}
 \label{tab:sim_results_bestfit}
\end{table}

We conclude that the statistical methods yield important evidence
against power-law-like distributions for our dataset, thus hinting
at the presence of a turnover in the ILF at absolute magnitudes
between $-8.0$ and $-9.5$ mag in the $U, B$, and $V$ band. In the $I$
band, the different models seem to fit the data with the same quality. Note that
even an ILF that is flat to the right of the peak of its best-fit
Gaussian would predict too many faint clusters, because even a
Gaussian distribution predicts rather too many than too few faint
clusters. However, such a model is at the borderline between any
possible presence and absence of a turnover in the intrinsic
distribution. In particular, a rising ILF such as a power law is even
less likely.

\subsection{Cluster ages, cluster masses and infant mortality}
\label{sec:ages}

We determined ages from our broad-band photometry using the AnalySED
algorithm, described in \citet{2004MNRAS.347..196A}. Both the best
ages and representative 1$\sigma$ uncertainty ranges are shown in
Fig. \ref{fig:age_distr}. Combining the average age of our full
cluster sample of $\approx$25 Myr and solar metallicity with the
parameters in Table \ref{tab:sim_results_bestfit} yields the following
parameter estimates for the MF: $\log(M_{\rm TO} [{\rm M}_\odot])=4.2$
and $\sigma_M=0.85$ dex. Compared to the values for the Milky Way (see
e.g.  \citealt{1998gcs..book.....A}, $\log(M_{\rm median} [{\rm
M}_\odot])=4.9$, $\log(M_{\rm mean} [{\rm M}_\odot])=5.3$,
$\sigma_M=0.49$ dex), this turn-over appears to be shifted to lower
masses, and to be broader. Note that we prefer to apply our
statistical treatment to the luminosities rather than to the
(physically more relevant) cluster masses, because the crucial
completeness determination can be performed accurately only for the
(observed) magnitudes and not for the (derived) masses.

Observations of increasing numbers of interacting and starburst
galaxies show a significantly larger number of young ($\la 10-30$ Myr)
star clusters than expected from a simple extrapolation of the cluster
numbers at older ages, under the assumption of a roughly constant star
cluster formation rate over the host galaxy's history, and taking into
account the observational completeness limits as well as the effects
of sample binning (see \citealt{grijs_china} for an in-depth
review). The current consensus is that an initial fast ($\la 10-30$
Myr) disruption mechanism can effectively remove up to 90 per cent of
the youngest, short-lived clusters from a given cluster
population. This process has been coined cluster ``infant mortality''
(\citealt{2004ASPC..322..419W}).

Taking the best ages in Fig. \ref{fig:age_distr} at face value, the
time-scale of infant mortality appears to be around 18-25 Myr
(depending on the binning in age), in rough agreement with previous
claims (see e.g.
\citealt{2004ASPC..322..419W,2005ApJ...631L.133F,2005A&A...443...41M};
see also \citealt{2006astro.ph.11055W} for a presentation of earlier
results), but perhaps somewhat shorter than theoretical predictions by
\citet{2006MNRAS.373..752G}, possibly due to the high galactic
background and/or the violent environment. Assuming a constant cluster
formation rate during the last $\lta$ 100 Myr the infant mortality
rate is of order 60 per cent, i.e. $\approx$40 per cent of the newly
formed clusters survive the first $\approx$25 Myr.

\begin{figure}%[h]
 \vspace{0.5cm}
 \begin{center}
 \vspace{-1.0cm}
 \includegraphics[angle=270,width=1.00\columnwidth]{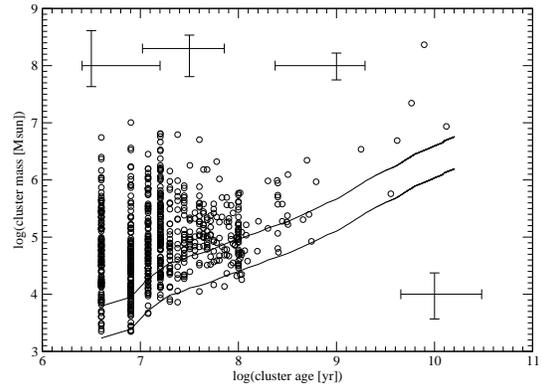}
 \end{center} 
 \vspace{-0.3cm}

 \caption{Age-vs.-mass relation for the full cluster sample. In the 
 lower-right corner mean error bars for all clusters are shown. The upper error
 crosses give the mean uncertainties for clusters with ages $<$ 10 Myr, between
 10 and 100 Myr and $>$ 100 Myr, respectively. The solid lines represent the
 fading lines, as given by the {\sc galev} models (solar metallicity, see
 \citealt{2003A&A...401.1063A}) and the completeness limits determined in
 Section \ref{sec:completeness}. Both lines correspond to the completeness
 curves for artificial clusters with 1 pixel size. Upper line: DISC region.
 Lower line: OVERLAP region.}

 \label{fig:age_distr} 
\end{figure}

\subsection{Investigating subsets of the full cluster sample}
\label{sec:subsets}

In this section we investigate the existence of a turnover in the LFs
of subsets of the full cluster sample, using our newly developed
statistics tools (Section \ref{subsection:statmodel}). We limit these
analyses to the $V$-band magnitudes.

First, we analysed the photometric data not corrected for
size-dependent ACs, in order to exclude the possibility of
calibration errors from the sizes/AC algorithm. The data for the
$V$ band are shown in Table \ref{tab:sim_results_subsets}.

In addition, we study three subsamples in cluster size (radius)
and several subsamples in cluster age.

Two of the three size bins were chosen to be centred around the
cluster sizes we studied the completeness functions for previously.
The third bin contains all clusters larger than in the first two bins.

First, we chose three age bins roughly equally spaced in
log(age/yr). In addition, we investigated subsamples which are (or are
not) expected to be affected by infant mortality: following the
theoretical work by \citet{2006MNRAS.373..752G} we divided the cluster
sample in clusters younger and older than 50 Myr. As a second test,
based on the observed age distribution of our Antennae cluster sample,
we divided the cluster sample into clusters younger and older than 25
Myr, as this seems to be the disruption time we see in our cluster
sample.

As can be seen from Table \ref{tab:sim_results_subsets}, all but one
of the subsets considered confirm the findings for the full data
set. The only exception involves clusters with ages $>$ 100 Myr, but
these subsets consist of 23/42 clusters, too few to base unambiguous
statistical conclusions on.

No significant difference is seen for the subsets during or after the
infant mortality phase.
\begin{table}
 \begin{center}

 \caption{Results of the Likelihood Ratio tests for several subsets,
 performed for the $V$ band. For clarity, the difference of the
 negative log-likelihood values is given instead of the likelihood
 ratios. Hence, negative values indicate that the null hypothesis
 (i.e. power law) is superior. The last column gives the probability
 of error for discarding an intrinsic power law (hence small values
 represent the inferiority of the power law). The {\it italic} values
 are for clusters in the size range FWHM=0.5-2.36 pixel
 $\rightarrow$ $r_{1/2} \simeq$ 5-25 pc.}

 \begin{tabular}{|l| r r r r r r|}
 \hline
 & & \multicolumn{3}{|c|}{\textbf{Quantiles}} & & \\
Subset & \# & 50\% & 99\% & 100\% & {\bf Obs.} & Prob.\\
 \hline
 & & \multicolumn{3}{|c|}{\textbf{Without ACs}} & & \\
 \hline
 $V$ band & 752 & -0.8 & 4.9 & 7.6 & {\bf 20.2} & 0.000 \\
 \hline
 {\it $V$ band} & 365 & -0.2 & 5.7 & 11.4 & {\bf 23.1} & 0.000 \\
 \hline
 & & \multicolumn{3}{|c|}{\textbf{Radius bins}} & & \\
 \hline
 $<$ 1.5 pix & 275 & -0.2 & 6.2 & 8.0 & {\bf 27.5} & 0.000 \\ 
 $[$1.5:2.5$]$ pix & 104 & -0.6 & 2.0 & 3.7 & {\bf 2.6} & 0.004 \\ 
 $>$ 2.5 pix & 373 & -8.0 & -0.4 & 1.1 & {\bf 24.4} & 0.000 \\ 
 \hline
 & & \multicolumn{3}{|c|}{\textbf{Age bins}} & & \\
 \hline
$\le$ 12 Myr & 382 & -3.9 & 2.4 & 9.4 & {\bf 4.8} & 0.001 \\
$[$16:100$]$ Myr & 328 & -2.2 & 3.0 & 6.9 & {\bf 8.7} & 0.000 \\
$>$ 100 Myr & 42 & -0.6 & 2.2 & 3.5 & {\bf 0.4} & 0.118 \\
$<$ 25 Myr & 541 & -5.4 & 1.9 & 5.7 & {\bf 5.0} &  0.001\\
$>$ 25 Myr & 211 & -0.8 & 4.2 & 7.5 & {\bf 13.5} & 0.000\\
$<$ 50 Myr & 627 & -5.7 & 2.0 & 7.5 & {\bf 10.6} & 0.000\\
$>$ 50 Myr & 125 & -0.4 & 3.8 & 6.6 & {\bf 7.2} &  0.000\\
 \hline
{\it $\le$ 12 Myr} & 179 & -0.6 & 4.2 & 6.8 & {\bf 9.1} & 0.000 \\
{\it  $[$16:100$]$ Myr} & 163 & -0.4 & 4.3 & 6.2 & {\bf 20.4} & 0.000 \\
{\it  $>$ 100 Myr} & 23 & -1.6 & 1.8 & 3.4 & {\bf 0.8} & 0.054  \\
{\it  $<$ 25 Myr} & 254 & -0.9 & 4.5 & 7.7 & {\bf 14.3} & 0.000\\
{\it  $>$ 25 Myr} & 111 &  0.5 & 5.0 & 6.5 & {\bf 20.1} & 0.000\\
{\it  $<$ 50 Myr} & 291 & -0.8 & 4.8 & 6.6 & {\bf 19.8} & 0.000\\
{\it  $>$ 50 Myr} & 74  & -2.7 & 3.4 & 8.2 & {\bf 6.5}  & 0.002\\
\hline
 \end{tabular}
 \label{tab:sim_results_subsets}
 \end{center}
\end{table}

\subsection{Investigating the effects of artificial decrease of
completeness}

We performed tests with the completeness functions artificially
degraded, using offsets of 0.5 mag, 1 mag and 2 mag. The results are
shown in Tables \ref{tab:sim_results_subsets_shift} and
\ref{tab:sim_results_bestfit_shift} and in Fig.
\ref{fig:distr_fit_shifted}.

As one might expect, the resulting best-fit model LFs become more and
more skewed towards the bright end with increasing offsets, becoming
increasingly unlike the observed distribution. For the full cluster
sample, the power-law ILF becomes increasingly superior with
increased offsets, as one might expect. For the sample containing
only the ``small' clusters, the superiority of the Gaussian is
retained even by degrading the completeness.  This follows naturally
from intrinsic selection effects in which larger clusters of a given
brightness will more easily be missed in our sample selection than
smaller clusters, owing to their lower surface brightnesses.

For the full cluster sample the peak of the Gaussian distribution
shifts strongly towards fainter magnitudes with increasing offset,
i.e. in the range where the completeness functions are significantly
greater than 0 both tested distributions resemble power laws. This
effect is not as significant for the cluster sample with the narrower
size range, again because of the intrinsic selection effects referred
to in this respect.

\begin{table}
 \begin{center}

 \caption{Results of the Likelihood Ratio tests for either the full
 sample or the subset with half-light radii between $\simeq$ 5 and 25
 pc, for different artificial shifts in the completeness functions,
 performed for the $V$ band. For clarity, the difference of the
 negative log-likelihood values is given instead of the likelihood
 ratios. Hence, negative values indicate that the null hypothesis
 (i.e.  power law) is superior. The last column gives the probability
 of error for discarding an intrinsic power law (hence small values
 represent the inferiority of the power law). These values relate to
 Fig. \ref{fig:distr_fit_shifted}.}

 \begin{tabular}{|l| r r r r r r|}
 \hline
 & & \multicolumn{3}{|c|}{\textbf{Quantiles}} & & \\
Subset & \# & 50\% & 99\% & 100\% & {\bf Obs.} & Prob.\\
 \hline
 & \multicolumn{6}{|c|}{\textbf{Shift by 0.5 mag, all clusters}} \\
 \hline
 $V$ band & 752 & -1.9 & 4.0 & 5.8 & {\bf 1.3} & 0.072 \\
 \hline
 & \multicolumn{6}{|c|}{\textbf{Shift by 1.0 mag, all clusters}} \\
 \hline
 $V$ band & 752 & -1.1 & 3.0 & 6.1 & {\bf -3.3} & 0.971 \\
 \hline
 & \multicolumn{6}{|c|}{\textbf{Shift by 2.0 mag, all clusters}} \\
 \hline
 $V$ band & 752 & -3.9 & 0.6 & 16.3 & {\bf -11.2} & 1.000 \\
 \hline
 & \multicolumn{6}{|c|}{\textbf{Shift by 0.5 mag, small clusters}} \\
 \hline
 $V$ band & 365 & -0.1 & 5.3 & 7.0 & {\bf 17.9} & 0.000 \\
 \hline
 & \multicolumn{6}{|c|}{\textbf{Shift by 1.0 mag, small clusters}} \\
 \hline
 $V$ band & 365 & 0.2 & 5.5 & 9.8 & {\bf 15.6} & 0.000 \\
 \hline
 & \multicolumn{6}{|c|}{\textbf{Shift by 2.0 mag, small clusters}} \\
 \hline
 $V$ band & 365 & 1.0 & 6.3 & 8.6 & {\bf 21.6} & 0.000 \\
 \hline
 \end{tabular}
 \label{tab:sim_results_subsets_shift}
 \end{center}
\end{table}

\begin{table}

 \caption{Best-fitting parameters for the Gaussian model, corresponding
 to Fig. \ref{fig:distr_fit_shifted} and Table
 \ref{tab:sim_results_subsets_shift}.}

 \begin{tabular}{|l| r c r |}
 \hline
 & \multicolumn{2}{|c|}{\textbf{Gaussian}} & \\
Filter & $\mu$ & 95 percentile range in $\mu$ & $\sigma$ \\
 \hline
 & \multicolumn{2}{|c|}{\textbf{Shift by 0.5 mag, all clusters}} & \\
 \hline
 $V$ & -6.8 & $[$-7.5:-5.9$]$ & 2.5 \\
 \hline
 & \multicolumn{2}{|c|}{\textbf{Shift by 1.0 mag, all clusters}} & \\
 \hline
 $V$ & -4.4 & $[$-5.6:-2.6$]$ & 2.8 \\
 \hline
 & \multicolumn{2}{|c|}{\textbf{Shift by 2.0 mag, all clusters}} & \\
 \hline
 $V$ & -2.0 & $[$-3.7:+0.7$]$ & 2.7 \\
 \hline
 & \multicolumn{2}{|c|}{\textbf{Shift by 0.5 mag, small clusters}} & \\
 \hline
 $V$ & -7.8 & $[$-8.1:-7.3$]$ & 1.5 \\
 \hline
 & \multicolumn{2}{|c|}{\textbf{Shift by 1.0 mag, small clusters}} & \\
 \hline
 $V$ & -7.4 & $[$-7.8:-6.8$]$ & 1.5 \\
 \hline
 & \multicolumn{2}{|c|}{\textbf{Shift by 2.0 mag, small clusters}} & \\
 \hline
 $V$ & -7.0 & $[$-7.5:-6.4$]$ & 1.3 \\
 \hline
 \end{tabular}
 \label{tab:sim_results_bestfit_shift}
\end{table}

\begin{figure*} 
\vspace{-0.4cm}
 \begin{tabular}{lcc}
 \parbox{6.5cm}{\hspace{3.8cm}Gaussian} & Power Law
 \\
 \parbox{2.5cm}{all clusters}
 \parbox{6cm}{\includegraphics{figs/All_AC_0.5_10_V-Band--Median_Fitted--Gauss.eps}}&
 \parbox{6cm}{\includegraphics{figs/All_AC_0.5_10_V-Band--Median_Fitted--Power-Law.eps}}
 \\
 \parbox{2.5cm}{all clusters\\1mag shift}
 \parbox{6cm}{\includegraphics{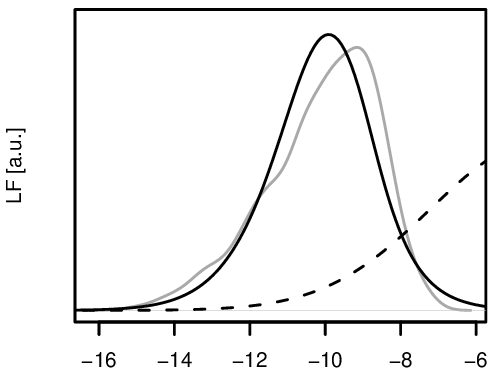}}&
 \parbox{6cm}{\includegraphics{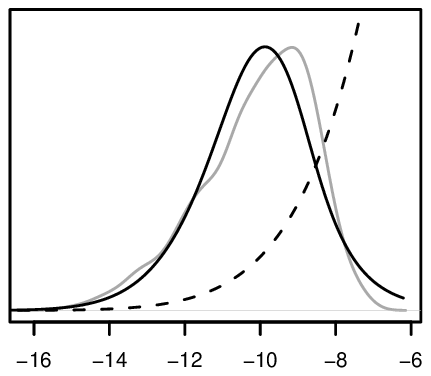}}
 \\ 
 \parbox{2.5cm}{all clusters\\2mag shift}
 \parbox{6cm}{\includegraphics{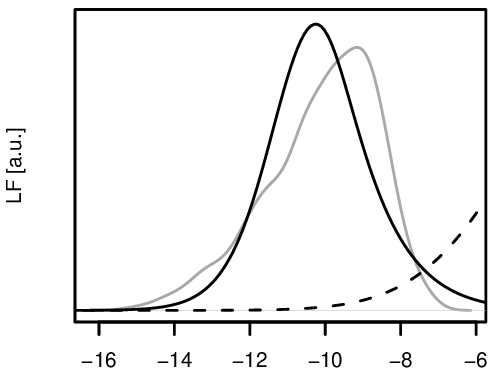}}&
 \parbox{6cm}{\includegraphics{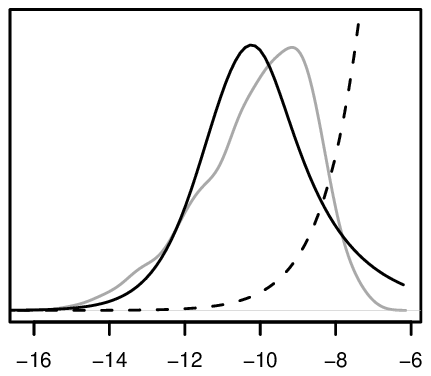}}
 \\ 
 \parbox{2.5cm}{small clusters}
 \parbox{6cm}{\includegraphics{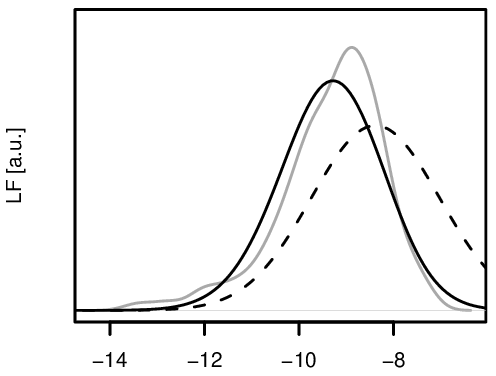}}&
 \parbox{6cm}{\includegraphics{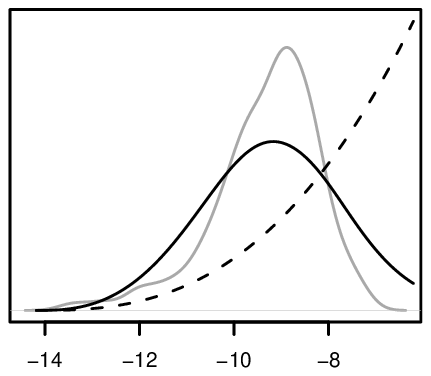}}
 \\
 \parbox{2.5cm}{small clusters\\1mag shift}
 \parbox{6cm}{\includegraphics{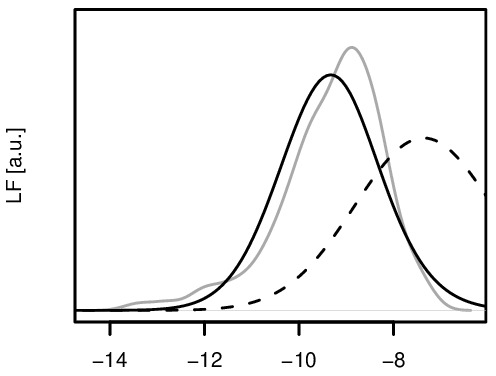}}&
 \parbox{6cm}{\includegraphics{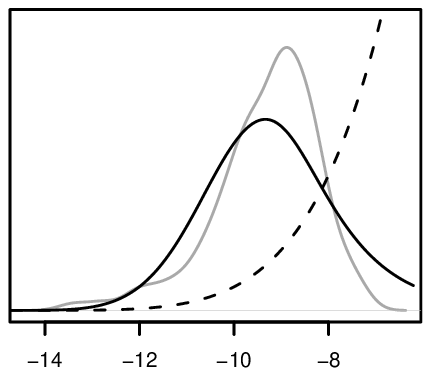}}
 \\
 \parbox{2.5cm}{small clusters\\2mag shift}
 \parbox{6cm}{\includegraphics{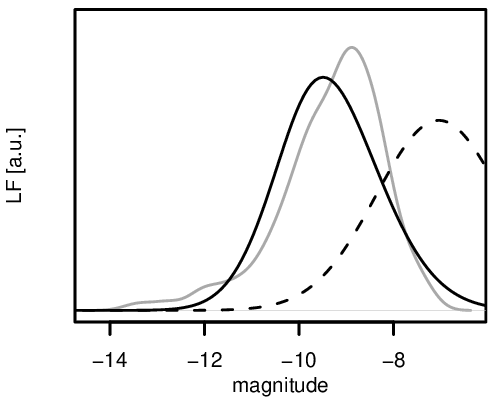}}&
 \parbox{6cm}{\includegraphics{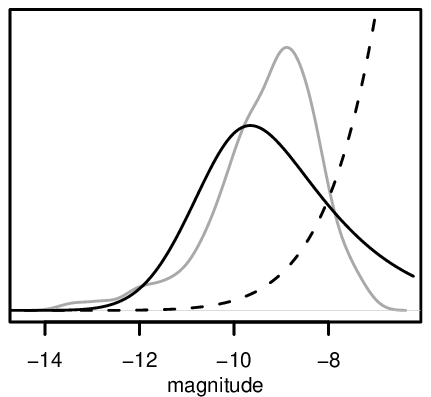}}
 \end{tabular}

 \caption{\label{fig:Model_fits_shift} Fits of the Gaussian and the
 power-law model for the differently shifted completeness curves, 
 $V$ band only. The figure coding is as in Fig. \ref{fig:Model_fits}.
 }
 \label{fig:distr_fit_shifted}
\end{figure*}

\section{Discussion}
\label{sec:discussion}

\subsection{Star Cluster Systems and their progenitors: LFs and MFs}

The mass spectra of the SC progenitors, GMCs and GMC cores, in nearby
galaxies are found to be of power law-type, with a slope of $-2.0 \lta
\alpha \lta -1.5$ over a wide range of cloud and core masses
(\citealt{1987ApJ...319..730S,1998A&A...329..249K,1998A&AS..133..337Z,
2005PASP..117.1403R,2006ApJ...641..389R}), possibly with the exception
of clouds in the outer parts of the Milky Way and in M33 (see
\citealt{2005PASP..117.1403R}), which might be represented by a
steeper power law. All these galaxies, however, are neither currently
interacting nor experiencing a strong starburst.

A first attempt to assess the molecular cloud mass spectrum in the
Antennae galaxies, by \citet{2003ApJ...599.1049W}, revealed a power
law with ${\rm \alpha \sim -1.4}$ but remained limited to GMC masses
above $5 \times 10^6$ M$_\odot$. Resolution of lower-mass molecular
clouds, observations of molecular cloud cores and the determination of
the cloud MF -- {\sl Is it different in massive gas-rich mergers
compared to non-interacting galaxies?} -- have to await commissioning
of ALMA.

For old GC systems both LFs and MFs are of Gaussian shape: the LFs
show a Gaussian turnover magnitude at $M_V = -7.3$ mag and a Gaussian
$\sigma = 1.2$ mag. As GCs seem to be roughly coeval, at least within
the associated uncertainties, with ages of 10--12 Gyr, the peaked
luminosity distribution can be converted into a mass distribution. The
corresponding peak mass occurs at $\sim 2 \times 10^5$ M$_\odot$. YSCs
in the local Universe show LFs that appear generally characterised by
power laws, at least towards the upper mass limit. It is not clear,
however, whether or not the distinction between YSCs and GCs is
meaningful for all galaxies, and whether it reflects an intrinsic
difference in the nature of these clusters (i.e. their formation
mechanisms) or originates from a initially universal continuous
cluster distribution re-shaped by secular evolution and destruction of
the lower-mass SCs.

\citet{2004A&A...416..537L} reports the detection of so-called super
star clusters (SSCs) in a number of seemingly undisturbed, quiescently
star-forming face-on spirals. These SSCs are clearly very bright and
very young; at least some of them have been shown to be very massive,
too (\citealt{2004AJ....128.2295L}). With masses around $2 \times
10^5$ to $1.5 \times 10^6$ M$_\odot$ and small radii (half-light radii
between 3.0 and 5.2 pc; see \citealt{2004AJ....128.2295L}), they
resemble young GCs, although ongoing GC formation is not expected in
these non-interacting, non-starbursting, just normally star-forming
galaxies (see also Section \ref{sec:sfe}). If these massive SSCs are not
very rare exceptions, we would have to investigate where the
descendants are of all those SSCs that presumably formed earlier on,
i.e. the intermediate-age GCs in those spirals, or whether we are
witnessing a very special epoch in the star-formation history of those
spirals.

Starburst and merging galaxies are observed to harbour rich
systems of YSCs formed during the starburst event. Merger remnants
with post-starburst signatures also reveal SC systems with ages of up
to 1--3 Gyr (e.g., \citealt{2003ApJ...583L..17D,2004ApJ...613L.121G}),
indicating that at least some fraction of these SCs formed during the
burst survived much longer than most of the open clusters in the Milky
Way.

The universality or non-universality (and possibly the environmental
dependence) of the LFs of SC systems in quiescently star-forming
galaxies compared to merger-induced starbursts provides important
input for models of star and star cluster formation.

\subsection{SC formation and star formation efficiencies}
\label{sec:sfe}

The lifetime of a SC depends on its mass (and stellar initial mass
function), its initial degree of binding, and on its environment.
While apparently all star formation is SC formation at the stage of
embedded clusters in the Milky Way, most of the clusters are already
unbound when they emerge from their dust cocoon
(\citealt{2003ARA&A..41...57L}). Hydrodynamical cluster formation
modelling has shown that very high SFEs are required to form a massive
cluster that is long-term stable. The formation of a typical GC,
sufficiently massive and strongly bound to survive for a Hubble time,
requires a SFE of order 30 per cent or higher
(\citealt{1995ApJ...440..666B,1997ApJ...480..235E, 2006MNRAS.369L...9B}),
i.e. at least an order of magnitude higher than SFEs in normal spiral
and irregular galaxies, or in dwarf galaxy starbursts (cf.
\citealt{1995A&A...303...41K,2002A&A...385..412M}). SFEs as high as 30
per cent or more, however, are observed in global and nuclear
starbursts triggered by massive gas-rich mergers, such as NGC 7252,
and ultra-luminous infrared galaxies (ULIRGs;
\citealt{1994A&A...285..775F,2004ApJ...606..271G}) and led to the idea
that GCs might form in these events and not only in the early Universe
(\citealt{1995A&A...300...58F,2002IAUS..207..630S,2003IAUJD..11E..44S}).

In addition, observations of a gap in cluster ages but not in field
star ages in the LMC (\citealt{2001AJ....122..842R}) suggest that SC
formation there took place in stages of enhanced star formation only,
possibly related to close encounters of the LMC with the Milky Way
and/or the SMC. Similarly, star cluster formation in M51 is found to
have been significantly enhanced during the last close encounter with
its companion NGC 5194 (\citealt{2005A&A...431..905B}), once again
when the overall star formation and the SFE are expected to be
enhanced.

Although little is known about molecular cloud {\sl structures} in
interacting and merging galaxies, a clear difference does exist in the
molecular gas {\sl content} between normal star-forming galaxies and
ULIRGs.  We know that while the CO(1-0) line traces molecular gas at
densities $n \geq 100 {\rm cm}^{-3}$, the HCN(1-0) and CS (1-0) lines
trace gas at densities $n \geq 10^4 {\rm cm}^{-3}$ and $n \geq 10^5
{\rm cm}^{-3}$, respectively.

Submillimetre observations show that for quiescently star-forming
galaxies only a small fraction $(\sim 0.1 - 3$ per cent) of all their
(CO) molecular gas is at the high densities of molecular cloud cores,
as traced by HCN or CS, i.e. $L({\rm HCN})/L({\rm CO}) \sim 0.001 -
0.03$. In contrast, ULIRGs show that, averaged over volumes of ($10 -
300$ pc)$^3$, the ratio $M({\rm HCN})/M({\rm CO})$ can reach up to
$0.3 - 1$ (\citealt{2004ApJ...606..271G}), i.e. the molecular cloud
structure must be very different from the case of quiescently
star-forming galaxies -- to the point that it becomes very difficult
to imagine much internal structure at all, if essentially all the
molecular gas is at molecular cloud core densities.

The mass ratio of high-density gas to the total gas mass (as traced by
the $L({\rm HCN})/L({\rm CO})$ ratio) in ULIRGs is higher by a factor
of $\sim 5-10$ when compared with normal star-forming galaxies
(\citealt{2004ApJ...606..271G}). Since the mass in high-density gas
correlates linearly with the star-formation rate (SFR; as measured by
the far-infrared flux; \citealt{2004ApJ...606..271G}), ULIRGs
therefore also possess a star formation per unit total gas mass higher
by a factor $\sim 5-10$ compared to normal SF galaxies.

The \citet{1959ApJ...129..243S} law relating the surface densities of
the SFR and of hydrogen gas (H{\sc i} from direct observations and
H$_2$ as traced by CO) by $\Sigma_{\rm SFR} \propto \Sigma_{\rm
gas}^n$ with $n \sim 1.4$ is valid from quiescently star-forming
spirals all the way through to ULIRGs, probably the most actively
star-forming galaxies in existence today
(\citealt{1998ApJ...498..541K}), over almost 5 orders of magnitude in
gas surface density and 6 orders of magnitude in SFR density. However,
if this relation is expressed in terms of high-density gas, as traced
by HCN or CS, instead of the low-density gas traced by CO, it becomes
linear, indicating that star formation is more fundamentally governed
by the content of {\sl high density} gas, not the {\sl overall} gas
content (see \citealt{1992ApJ...387L..55S,2004ApJ...606..271G}). The
non-linearity in the \citet{1998ApJ...498..541K} law might then arise
from an environment-dependent time scale and/or efficiency to
transform H{\sc i} into H$_2$ and the low-density gas into
high-density gas. This is an important issue to consider in
hydrodynamical modelling of galaxies and galaxy mergers, which then
needs to account for a multi-phase ISM and include a careful
description of phase transitions, star formation and feedback
processes.

Star formation in normal galaxies, spirals and irregulars, is thought
to occur through the collapse of molecular clouds, whereby the mass
spectrum apparently remains self-similar from molecular clouds through
molecular cloud cores all the way to the mass spectrum of open star
clusters, all of which are power laws with $m \sim -1.5$ to $-2$
(\citealt{2003ARA&A..41...57L}, see \citealt{1997ApJ...480..235E} for
a theoretical foundation), implying a constant SFE. However,
observations show considerable SFE scatter between different clouds
and clusters (see e.g.
\citealt{1997ApJ...488..286L,1999ApJ...518..760K,2003ARA&A..41...57L}),
although this quantity is naturally very difficult to measure. As the
SFE is linked strongly non-linearly with the surviving bound fraction
of the cluster
(\citealt{1984ApJ...285..141L,2001MNRAS.323..988G,2005ApJ...630..879F}),
deviations from this self-similarity are not coming as a
surprise. Their impact on the form of the cluster LF (and MF) is
studied extensively in \citet{parmentier06}, where it is shown to
explain the occurrence of a turnover in the cluster LF derived from a
power-law cloud LF (or MF). However, this whole issue is currently
under vigorous debate (see also below).

In interacting galaxies, the frequency of molecular cloud collisions
is expected to increase strongly and this will considerably enhance
star formation. Moreover, molecular clouds get shock-compressed by
external pressure (recently verified observationally for the Antennae
galaxies by \citealt{2005A&A...433L..17H}, although
\citealt{2005AJ....130.2104W} report the absence of shock-heated gas
close to their sample of YSCs), grow denser and more massive, and this
process can drive up the SFE very efficiently
(\citealt{1992ApJ...387..152J,2004MNRAS.350..798B}).
\citet{1992ApJ...400..476J,1996ApJ...473..797J} have shown that a
relatively small increase in the external ambient pressure to values 3
-- 4 times the internal pressure within the molecular clouds in the
undisturbed galaxy can drive SFEs up to 70 -- 90 per cent.

\subsection{Evolution of SC systems}

Based on the observed power-law LFs of YSCs by
\citet{1984AJ.....89.1822V} for the Milky Way and
\citet{2003AJ....126.1836H} for the LMC, a power-law LF is usually
also adopted for YSC systems in interacting galaxies (see
\citealt{2003MNRAS.343.1285D} for a recent compilation; also
\citealt{2001ApJ...561..727Z,1999AJ....118..752Z,2001AJ....121..768D}),
although in most cases these LFs cannot be traced to below the
expected turnover magnitude, if there were one. In addition, in such
violently star-forming systems the stellar contamination of the sample
is more difficult to deal with, and the determination of observational
completeness fractions is difficult because of strongly variable
backgrounds. Non-negligible age differences among the YSCs lead to
distortions of the LFs with respect to their underlying, intrinsic MFs
(\citealt{1995Natur.375..742M,1999A&A...342L..25F}).

Approaching this issue from the opposite point of view, it is usually
argued that the underlying MF is most likely a power-law, based on
observations of the GMC MFs in nearby normal galaxies (e.g.,
\citealt{1987ApJ...319..730S,1997ApJ...480..235E,1998A&A...329..249K,
1998A&AS..133..337Z,2005PASP..117.1403R,2006ApJ...641..389R}). This
argument has two drawbacks, however:

\begin{enumerate}

\item It assumes that the cluster mass correlates directly with the
mass of the precursor GMC, and hence must be subject to a constant
SFE. As indirect measure, a correlation between mass and radius would
then be expected.  The presence of a mass-radius relation for GMCs and
its absence for clusters (\citealt{2001AJ....122.1888A}), however,
casts doubt on the validity of this assumption. Present studies on
these mass-radius relations are limited to nearby and therefore only
relatively undisturbed galaxies, however. The impact of the observed
cloud-cloud scatter in SFEs on the cluster LF/MF was recently studied
by \citet{parmentier06}. Here, it was shown that this scatter can
transform a power-law MF of GMCs into a bell-shaped star cluster MF.

\item It assumes that the GMC MFs in interacting galaxies are similar
to those in nearby normal galaxies. However, as recently shown
(\citealt{2004ApJ...606..271G}), of relevance for star formation is
not the mass of the GMCs, but the mass of the dense cores embedded
within these GMCs. The fraction of gas in these cores compared to the
total gas mass is substantially higher in violently interacting and
star-forming galaxies, compared to quiescently star-forming isolated
galaxies (\citealt{2004ApJ...606..271G}). The MFs of both GMCs and of
their cores in the nearest interacting and starburst galaxies,
e.g. such as in the Antennae galaxies, will be determined only to
sufficiently low masses\footnote{Following \citet{2003ApJ...599.1049W}
and the ALMA sensitivity calculator\\
(http://www.eso.org/projects/alma/science/bin/sensitivity.html), we
roughly estimate limiting GMC masses to be observed in the Antennae
galaxies by ALMA around $6 \times 10^4$ M$_\odot$ (1 hr observation)
to $1 \times 10^4$ M$_\odot$ (30 hr observation).} when the next
generation (sub)millimetre observatory ALMA becomes operational in the
next decade. Until then, the initial cluster MF will likely be the
best proxy, but see (i) above.
\end{enumerate}

It has hitherto remained unclear whether the difference in shape between
the power-law LFs of young SCs (but see
\citealt{1998A&A...336...83F,1999A&A...342L..25F},
\citealt{2003ApJ...583L..17D} and \citealt{2004ApJ...613L.121G} for the
Gaussian LFs in the Antennae galaxies, M82B and NGC 1316, and the new
results by \citealt{2006MNRAS.366..295D} on the LMC SC system) and old GC
systems is caused by differences in the nature and formation of the two
types of clusters, or whether the power law of young systems is secularly
transformed into the Gaussian distribution of old GC systems by selective
destruction effects. Models for the evolution of SC systems in
galactic potentials have naturally obtained a Gaussian shape from an
initial power law by selectively destroying low-mass clusters
(\citealt{2001ApJ...561..751F}). Detailed dynamical studies that also
include dynamical friction as an important disruption mechanism for massive
clusters (\citealt{2000MNRAS.318..841V}) show that although it is possible
to obtain a final Gaussian distribution from an initial power-law,
significant fine-tuning of the model parameters is required to obtain the
observed GC mass function. In contrast, an initial  Gaussian shape similar
to that observed today is conserved for a wide range of assumptions despite
the disruption of more than 50 per cent of the original cluster population.
In addition, an initial Gaussian shape  with parameters differing from the
presently observed ones tend to evolve to a shape with the presently
observed parameters at GC ages (\citealt{2000MNRAS.318..841V}). It is not
yet possible to do this kind of modelling in the time-varying galactic
potentials of merging galaxies.  \citet{2005MNRAS.363..326P,parmentier06}
have shown that the initial cluster MF of the Galactic GC system very
probably was already either Gaussian in shape or described by a truncated
power law with a cut-off at the same position as the turnover of the
Gaussian distribution.  Similarly, \citet{2005MNRAS.364.1054D} show that
the {\sl initial} cluster MF in the post-starburst region B in M82 (shown
in \citealt{2003ApJ...583L..17D} to possess a {\sl present-day} cluster MF
of roughly Gaussian shape) was also most likely of Gaussian shape,
supported by plausibility arguments regarding the initial stellar density
in this region.

\subsection{Our results in this general framework}

Our findings represent the youngest star cluster system for which a
Gaussian-shaped LF fits statistically significantly better than a
power-law distribution. This is an extension of the results found
observationally by \citet{2004ApJ...613L.121G} ($\simeq$ 3 Gyr-old
cluster system in the merger remnant NGC 1316) and
\citet{2003ApJ...583L..17D} ($\simeq$ 1 Gyr-old cluster system in the
fossil starburst region B in M82). It supports the detailed models by
Vesperini (see \citealt{2000MNRAS.318..841V} for a Gaussian initial LF
and \citealt{2001MNRAS.322..247V} for the power-law case),
(\citealt{2005MNRAS.363..326P,parmentier06}) and
\citet{2005MNRAS.364.1054D}. In the framework of \citet{parmentier06}
the occurrence of the turnover at such early times is a natural
outcome of shaping the LF during the earliest stages of cluster
evolution, namely the phase of the removal of gas left over from
cluster formation.  \citet{2000MNRAS.318..841V} and
\citet{2005MNRAS.364.1054D} show the necessity for an initial or very
early Gaussian LF, which is in agreement with \citet{parmentier06}.

The proposed universality of the Gaussian {\sl shape} of SC LFs,
although not necessarily their (possibly environment-dependent)
parameters, has far-reaching consequences for studies of star and star
cluster formation. It opens the possibility to study the formation of
massive bound clusters (proto-GCs) at close reach, a process that was
previously thought to have taken place exclusively in the early
Universe.

\section{Summary}
\label{sec:summary}

In this paper we have studied the luminosity functions of the YSCs in
the Antennae galaxies (NGC 4038/39), the nearest ongoing merger of two
spiral galaxies. The merging process is accompanied by a strong burst
of star formation, and particularly star cluster formation. We
carefully select a sample of clearly extended sources, and hence
remove the otherwise strong contamination by bright stars within the
Antennae galaxies.

The complex background, caused by the interaction and the
contamination by bright stars, hampers the accurate determination of
the observational completeness. However, an accurately determined
completeness function is essential to determine whether or not the
intrinsic LF of the cluster system shows a turnover like the old GC
systems. We present a number of tests on the completeness functions,
and showed that it is essential to include {\sl all} cluster selection
criteria in the determination of the completeness function.

We attempted to fit the intrinsic LFs with the most widely used
models, i.e. a power-law distribution and a Gaussian distribution,
taking the completeness function and the photometric errors into
account. We find statistically significant evidence that the LFs of
this sample of clearly extended sources are best described by an
intrinsic Gaussian distribution. We determine the chance of error
(i.e. the probability that we erroneously discard a better-fitting
power-law distribution) using Monte Carlo-simulations, and find it to
be below 0.5 per cent.

Earlier claims of power-law LFs suffer most likely from the strong
stellar contamination and the difficulties in the completeness
determination. In addition, the statistical tools developed for
this study are beyond the sensitivity of any commonly used tools.

\section*{acknowledgements}

PA wishes to thank the Astronomical Institute, Academy of Sciences of
the Czech Republic in Prague, where early parts of this paper were
written, and especially Jan Palous for useful discussions. PA
acknowledges support from DFG grant FR 911/11-3. L. Boysen
acknowledges support from the Georg Lichtenberg programme ``Applied
Statistics \& Empirical Methods'' and DFG graduate programme 1023
``Identification in Mathematical Models''. We would like to thank
B. Whitmore and M. Fall for providing us with their non-published full
set of source magnitudes for our comparison. We would also like to
express our thanks to C. Wilson and R. Kennicutt for fast
clarification of a number of issues dealt with in this paper, and to
the referee for pointing at several details to strengthen the paper.

\bibliographystyle{mn2e}
\bibliography{references,references_add}

\begin{thebibliography}{}

\bibitem[\protect\citeauthoryear{{Anders}, {Bissantz}, {Fritze-v.~Alvensleben}
  \& {de Grijs}}{{Anders} et~al.}{004a}]{2004MNRAS.347..196A}
{Anders} P.,  {Bissantz} N.,  {Fritze-v.~Alvensleben} U.,    {de Grijs} R.,
  2004a, \mnras, 347, 196

\bibitem[\protect\citeauthoryear{{Anders}, {de Grijs}, {Fritze-v.~Alvensleben}
  \& {Bissantz}}{{Anders} et~al.}{004b}]{2004MNRAS.347...17A}
{Anders} P.,  {de Grijs} R.,  {Fritze-v.~Alvensleben} U.,    {Bissantz} N.,
  2004b, \mnras, 347, 17

\bibitem[\protect\citeauthoryear{{Anders} \& {Fritze-v.~Alvensleben}}{{Anders}
  \& {Fritze-v.~Alvensleben}}{2003}]{2003A&A...401.1063A}
{Anders} P.,  {Fritze-v.~Alvensleben} U.,  2003, \aap, 401, 1063

\bibitem[\protect\citeauthoryear{{Anders}, {Gieles} \& {de Grijs}}{{Anders}
  et~al.}{2006}]{2006A&A...451..375A}
{Anders} P.,  {Gieles} M.,    {de Grijs} R.,  2006, \aap, 451, 375

\bibitem[\protect\citeauthoryear{{Ashman}, {Conti} \& {Zepf}}{{Ashman}
  et~al.}{1995}]{1995AJ....110.1164A}
{Ashman} K.~M.,  {Conti} A.,    {Zepf} S.~E.,  1995, \aj, 110, 1164

\bibitem[\protect\citeauthoryear{{Ashman} \& {Zepf}}{{Ashman} \&
  {Zepf}}{1998}]{1998gcs..book.....A}
{Ashman} K.~M.,  {Zepf} S.~E.,  1998, {Globular Cluster Systems}.
{Cambridge astrophysics series ; 30}, {Cambridge University Press}, {Cambridge,
  U.~K.~; New York}

\bibitem[\protect\citeauthoryear{{Ashman} \& {Zepf}}{{Ashman} \&
  {Zepf}}{2001}]{2001AJ....122.1888A}
{Ashman} K.~M.,  {Zepf} S.~E.,  2001, \aj, 122, 1888

\bibitem[\protect\citeauthoryear{{Barnes}}{{Barnes}}{1988}]{1988ApJ...331..699%
B}
{Barnes} J.~E.,  1988, \apj, 331, 699

\bibitem[\protect\citeauthoryear{{Barnes}}{{Barnes}}{2004}]{2004MNRAS.350..798%
B}
{Barnes} J.~E.,  2004, \mnras, 350, 798

\bibitem[\protect\citeauthoryear{{Bastian}, {Emsellem}, {Kissler-Patig} \&
  {Maraston}}{{Bastian} et~al.}{2006}]{2006A&A...445..471B}
{Bastian} N.,  {Emsellem} E.,  {Kissler-Patig} M.,    {Maraston} C.,  2006,
  \aap, 445, 471

\bibitem[\protect\citeauthoryear{{Bastian}, {Gieles}, {Efremov} \&
  {Lamers}}{{Bastian} et~al.}{005b}]{2005A&A...443...79B}
{Bastian} N.,  {Gieles} M.,  {Efremov} Y.~N.,    {Lamers} H.~J.~G.~L.~M.,
  {2005b}, \aap, 443, 79

\bibitem[\protect\citeauthoryear{{Bastian}, {Gieles}, {Lamers}, {Scheepmaker}
  \& {de Grijs}}{{Bastian} et~al.}{005a}]{2005A&A...431..905B}
{Bastian} N.,  {Gieles} M.,  {Lamers} H.~J.~G.~L.~M.,  {Scheepmaker} R.~A.,
  {de Grijs} R.,  {2005a}, \aap, 431, 905

\bibitem[\protect\citeauthoryear{{Bastian} \& {Goodwin}}{{Bastian} \&
  {Goodwin}}{2006}]{2006MNRAS.369L...9B}
{Bastian} N.,  {Goodwin} S.~P.,  2006, \mnras, 369, L9

\bibitem[\protect\citeauthoryear{{Brown}, {Burkert} \& {Truran}}{{Brown}
  et~al.}{1995}]{1995ApJ...440..666B}
{Brown} J.~H.,  {Burkert} A.,    {Truran} J.~W.,  1995, \apj, 440, 666

\bibitem[\protect\citeauthoryear{{Bruzual} \& {Charlot}}{{Bruzual} \&
  {Charlot}}{2003}]{2003MNRAS.344.1000B}
{Bruzual} G.,  {Charlot} S.,  2003, \mnras, 344, 1000

\bibitem[\protect\citeauthoryear{{Cresci}, {Vanzi} \& {Sauvage}}{{Cresci}
  et~al.}{2005}]{2005A&A...433..447C}
{Cresci} G.,  {Vanzi} L.,    {Sauvage} M.,  2005, \aap, 433, 447

\bibitem[\protect\citeauthoryear{{de Grijs} \& {Anders}}{{de Grijs} \&
  {Anders}}{2006}]{2006MNRAS.366..295D}
{de Grijs} R.,  {Anders} P.,  2006, \mnras, 366, 295

\bibitem[\protect\citeauthoryear{{de Grijs}, {Anders}, {Bastian}, {Lynds},
  {Lamers} \& {O'Neil}}{{de Grijs} et~al.}{003b}]{2003MNRAS.343.1285D}
{de Grijs} R.,  {Anders} P.,  {Bastian} N.,  {Lynds} R.,  {Lamers}
  H.~J.~G.~L.~M.,    {O'Neil} E.~J.,  2003b, \mnras, 343, 1285

\bibitem[\protect\citeauthoryear{{de Grijs}, {Bastian} \& {Lamers}}{{de Grijs}
  et~al.}{003c}]{2003ApJ...583L..17D}
{de Grijs} R.,  {Bastian} N.,    {Lamers} H.~J.~G.~L.~M.,  2003c, \apjl, 583,
  L17

\bibitem[\protect\citeauthoryear{{de Grijs}, {Lee}, {Clemencia Mora Herrera},
  {Fritze-v.~Alvensleben} \& {Anders}}{{de Grijs}
  et~al.}{003a}]{2003NewA....8..155D}
{de Grijs} R.,  {Lee} J.~T.,  {Clemencia Mora Herrera} M.,
  {Fritze-v.~Alvensleben} U.,    {Anders} P.,  2003a, New Astronomy, 8, 155

\bibitem[\protect\citeauthoryear{{de Grijs}, {O'Connell} \& {Gallagher}
  III}{{de Grijs} et~al.}{2001}]{2001AJ....121..768D}
{de Grijs} R.,  {O'Connell} R.~W.,    {Gallagher} III J.~S.,  2001, \aj, 121,
  768

\bibitem[\protect\citeauthoryear{{de Grijs} \& {Parmentier}}{{de Grijs} \&
  {Parmentier}}{2007}]{grijs_china}
{de Grijs} R.,  {Parmentier} G.,  2007, submitted to ChJA\&A

\bibitem[\protect\citeauthoryear{{de Grijs}, {Parmentier} \& {Lamers}}{{de
  Grijs} et~al.}{2005}]{2005MNRAS.364.1054D}
{de Grijs} R.,  {Parmentier} G.,    {Lamers} H.~J.~G.~L.~M.,  2005, \mnras,
  364, 1054

\bibitem[\protect\citeauthoryear{{Elmegreen} \& {Efremov}}{{Elmegreen} \&
  {Efremov}}{1997}]{1997ApJ...480..235E}
{Elmegreen} B.~G.,  {Efremov} Y.~N.,  1997, \apj, 480, 235

\bibitem[\protect\citeauthoryear{{Elson} \& {Fall}}{{Elson} \&
  {Fall}}{1985}]{1985PASP...97..692E}
{Elson} R.~A.~W.,  {Fall} S.~M.,  1985, \pasp, 97, 692

\bibitem[\protect\citeauthoryear{{Elson}, {Fall} \& {Freeman}}{{Elson}
  et~al.}{1987}]{1987ApJ...323...54E}
{Elson} R.~A.~W.,  {Fall} S.~M.,    {Freeman} K.~C.,  1987, \apj, 323, 54

\bibitem[\protect\citeauthoryear{{Fabbiano}, {Krauss}, {Zezas}, {Rots} \&
  {Neff}}{{Fabbiano} et~al.}{2003}]{2003ApJ...598..272F}
{Fabbiano} G.,  {Krauss} M.,  {Zezas} A.,  {Rots} A.,    {Neff} S.,  2003,
  \apj, 598, 272

\bibitem[\protect\citeauthoryear{{Fall}, {Chandar} \& {Whitmore}}{{Fall}
  et~al.}{2005}]{2005ApJ...631L.133F}
{Fall} S.~M.,  {Chandar} R.,    {Whitmore} B.~C.,  2005, \apjl, 631, L133

\bibitem[\protect\citeauthoryear{{Fall} \& {Zhang}}{{Fall} \&
  {Zhang}}{2001}]{2001ApJ...561..751F}
{Fall} S.~M.,  {Zhang} Q.,  2001, \apj, 561, 751

\bibitem[\protect\citeauthoryear{{Fellhauer} \& {Kroupa}}{{Fellhauer} \&
  {Kroupa}}{2005}]{2005ApJ...630..879F}
{Fellhauer} M.,  {Kroupa} P.,  2005, \apj, 630, 879

\bibitem[\protect\citeauthoryear{{Forbes}}{{Forbes}}{1996}]{1996AJ....112..954%
F}
{Forbes} D.~A.,  1996, \aj, 112, 954

\bibitem[\protect\citeauthoryear{{Fritze-v.~Alvensleben}}{{Fritze-v.~Alvensleb%
en}}{1998}]{1998A&A...336...83F}
{Fritze-v.~Alvensleben} U.,  1998, \aap, 336, 83

\bibitem[\protect\citeauthoryear{{Fritze-v.~Alvensleben}}{{Fritze-v.~Alvensleb%
en}}{1999}]{1999A&A...342L..25F}
{Fritze-v.~Alvensleben} U.,  1999, \aap, 342, L25

\bibitem[\protect\citeauthoryear{{Fritze-v.~Alvensleben}}{{Fritze-v.~Alvensleb%
en}}{2004}]{2004A&A...414..515F}
{Fritze-v.~Alvensleben} U.,  2004, \aap, 414, 515

\bibitem[\protect\citeauthoryear{{Fritze-v.~Alvensleben} \&
  {Burkert}}{{Fritze-v.~Alvensleben} \& {Burkert}}{1995}]{1995A&A...300...58F}
{Fritze-v.~Alvensleben} U.,  {Burkert} A.,  1995, \aap, 300, 58

\bibitem[\protect\citeauthoryear{{Fritze-v.~Alvensleben} \&
  {Gerhard}}{{Fritze-v.~Alvensleben} \& {Gerhard}}{1994}]{1994A&A...285..775F}
{Fritze-v.~Alvensleben} U.,  {Gerhard} O.~E.,  1994, \aap, 285, 775

\bibitem[\protect\citeauthoryear{{Gao} \& {Solomon}}{{Gao} \&
  {Solomon}}{2004}]{2004ApJ...606..271G}
{Gao} Y.,  {Solomon} P.~M.,  2004, \apj, 606, 271

\bibitem[\protect\citeauthoryear{{Geyer} \& {Burkert}}{{Geyer} \&
  {Burkert}}{2001}]{2001MNRAS.323..988G}
{Geyer} M.~P.,  {Burkert} A.,  2001, \mnras, 323, 988

\bibitem[\protect\citeauthoryear{{Gieles}, {Larsen}, {Scheepmaker}, {Bastian},
  {Haas} \& {Lamers}}{{Gieles} et~al.}{2006}]{2006A&A...446L...9G}
{Gieles} M.,  {Larsen} S.~S.,  {Scheepmaker} R.~A.,  {Bastian} N.,  {Haas}
  M.~R.,    {Lamers} H.~J.~G.~L.~M.,  2006, \aap, 446, L9

\bibitem[\protect\citeauthoryear{{Goodwin} \& {Bastian}}{{Goodwin} \&
  {Bastian}}{2006}]{2006MNRAS.373..752G}
{Goodwin} S.~P.,  {Bastian} N.,  2006, \mnras, 373, 752

\bibitem[\protect\citeauthoryear{{Goudfrooij}, {Gilmore}, {Whitmore} \&
  {Schweizer}}{{Goudfrooij} et~al.}{2004}]{2004ApJ...613L.121G}
{Goudfrooij} P.,  {Gilmore} D.,  {Whitmore} B.~C.,    {Schweizer} F.,  2004,
  \apjl, 613, L121

\bibitem[\protect\citeauthoryear{{Haas}, {Chini} \& {Klaas}}{{Haas}
  et~al.}{2005}]{2005A&A...433L..17H}
{Haas} M.,  {Chini} R.,    {Klaas} U.,  2005, \aap, 433, L17

\bibitem[\protect\citeauthoryear{{Harris}}{{Harris}}{1991}]{1991ARA&A..29..543%
H}
{Harris} W.~E.,  1991, \araa, 29, 543

\bibitem[\protect\citeauthoryear{{Hibbard},  \& {et al.}}{{Hibbard}
  et~al.}{2005}]{2005ApJ...619L..87H}
{Hibbard} J.~E.,     {et al.} 2005, \apjl, 619, L87

\bibitem[\protect\citeauthoryear{{Hibbard}, {van der Hulst}, {Barnes} \&
  {Rich}}{{Hibbard} et~al.}{2001}]{2001AJ....122.2969H}
{Hibbard} J.~E.,  {van der Hulst} J.~M.,  {Barnes} J.~E.,    {Rich} R.~M.,
  2001, \aj, 122, 2969

\bibitem[\protect\citeauthoryear{{Hunter}, {Elmegreen}, {Dupuy} \&
  {Mortonson}}{{Hunter} et~al.}{2003}]{2003AJ....126.1836H}
{Hunter} D.~A.,  {Elmegreen} B.~G.,  {Dupuy} T.~J.,    {Mortonson} M.,  2003,
  \aj, 126, 1836

\bibitem[\protect\citeauthoryear{{Jog} \& {Das}}{{Jog} \&
  {Das}}{1992}]{1992ApJ...400..476J}
{Jog} C.~J.,  {Das} M.,  1992, \apj, 400, 476

\bibitem[\protect\citeauthoryear{{Jog} \& {Das}}{{Jog} \&
  {Das}}{1996}]{1996ApJ...473..797J}
{Jog} C.~J.,  {Das} M.,  1996, \apj, 473, 797

\bibitem[\protect\citeauthoryear{{Jog} \& {Solomon}}{{Jog} \&
  {Solomon}}{1992}]{1992ApJ...387..152J}
{Jog} C.~J.,  {Solomon} P.~M.,  1992, \apj, 387, 152

\bibitem[\protect\citeauthoryear{{Kassin}, {Frogel}, {Pogge}, {Tiede} \&
  {Sellgren}}{{Kassin} et~al.}{2003}]{2003AJ....126.1276K}
{Kassin} S.~A.,  {Frogel} J.~A.,  {Pogge} R.~W.,  {Tiede} G.~P.,    {Sellgren}
  K.,  2003, \aj, 126, 1276

\bibitem[\protect\citeauthoryear{{Kavelaars}, {Harris}, {Hanes}, {Hesser} \&
  {Pritchet}}{{Kavelaars} et~al.}{2000}]{2000ApJ...533..125K}
{Kavelaars} J.~J.,  {Harris} W.~E.,  {Hanes} D.~A.,  {Hesser} J.~E.,
  {Pritchet} C.~J.,  2000, \apj, 533, 125

\bibitem[\protect\citeauthoryear{{Kennicutt}
  Jr.}{{Kennicutt}}{1998}]{1998ApJ...498..541K}
{Kennicutt} Jr. R.~C.,  1998, \apj, 498, 541

\bibitem[\protect\citeauthoryear{{King}}{{King}}{1962}]{1962AJ.....67..471K}
{King} I.,  1962, \aj, 67, 471

\bibitem[\protect\citeauthoryear{{Knierman}, {Gallagher}, {Charlton},
  {Hunsberger}, {Whitmore}, {Kundu}, {Hibbard} \& {Zaritsky}}{{Knierman}
  et~al.}{2003}]{2003AJ....126.1227K}
{Knierman} K.~A.,  {Gallagher} S.~C.,  {Charlton} J.~C.,  {Hunsberger} S.~D.,
  {Whitmore} B.,  {Kundu} A.,  {Hibbard} J.~E.,    {Zaritsky} D.,  2003, \aj,
  126, 1227

\bibitem[\protect\citeauthoryear{{Koo}}{{Koo}}{1999}]{1999ApJ...518..760K}
{Koo} B.-C.,  1999, \apj, 518, 760

\bibitem[\protect\citeauthoryear{{Kramer}, {Stutzki}, {Rohrig} \&
  {Corneliussen}}{{Kramer} et~al.}{1998}]{1998A&A...329..249K}
{Kramer} C.,  {Stutzki} J.,  {Rohrig} R.,    {Corneliussen} U.,  1998, \aap,
  329, 249

\bibitem[\protect\citeauthoryear{{Krist} \& {Hook}}{{Krist} \&
  {Hook}}{2004}]{krist97}
{Krist} J.,  {Hook} R.,  2004, {The Tiny Tim User's Guide}.
STScI, Baltimore, 6.3 edn

\bibitem[\protect\citeauthoryear{{Krueger}, {Fritze-v.~Alvensleben} \&
  {Loose}}{{Krueger} et~al.}{1995}]{1995A&A...303...41K}
{Krueger} H.,  {Fritze-v.~Alvensleben} U.,    {Loose} H.-H.,  1995, \aap, 303,
  41

\bibitem[\protect\citeauthoryear{{Lada} \& {Lada}}{{Lada} \&
  {Lada}}{2003}]{2003ARA&A..41...57L}
{Lada} C.~J.,  {Lada} E.~A.,  2003, \araa, 41, 57

\bibitem[\protect\citeauthoryear{{Lada}, {Margulis} \& {Dearborn}}{{Lada}
  et~al.}{1984}]{1984ApJ...285..141L}
{Lada} C.~J.,  {Margulis} M.,    {Dearborn} D.,  1984, \apj, 285, 141

\bibitem[\protect\citeauthoryear{{Lada}, {Evans} II \& {Falgarone}}{{Lada}
  et~al.}{1997}]{1997ApJ...488..286L}
{Lada} E.~A.,  {Evans} II N.~J.,    {Falgarone} E.,  1997, \apj, 488, 286

\bibitem[\protect\citeauthoryear{{Larsen}}{{Larsen}}{1999}]{1999A&AS..139..393%
L}
{Larsen} S.~S.,  1999, \aaps, 139, 393

\bibitem[\protect\citeauthoryear{{Larsen}}{{Larsen}}{2004}]{2004A&A...416..537%
L}
{Larsen} S.~S.,  2004, \aap, 416, 537

\bibitem[\protect\citeauthoryear{{Larsen}, {Brodie} \& {Hunter}}{{Larsen}
  et~al.}{2004}]{2004AJ....128.2295L}
{Larsen} S.~S.,  {Brodie} J.~P.,    {Hunter} D.~A.,  2004, \aj, 128, 2295

\bibitem[\protect\citeauthoryear{{Lehmann}}{{Lehmann}}{1994}]{lehmann94}
{Lehmann} E.~L.,  1994, {Testing statistical hypotheses}.
Chapman \& Hall, New York

\bibitem[\protect\citeauthoryear{{Little} \& {Rubin}}{{Little} \&
  {Rubin}}{2002}]{little02}
{Little} R.~J.~A.,  {Rubin} D.~B.,  2002, {Statistical analysis with missing
  data}, second edn.
Wiley Series in Probability and Statistics, Wiley-Interscience [John Wiley \&
  Sons], Hoboken, NJ

\bibitem[\protect\citeauthoryear{{McLaughlin} \& {Pudritz}}{{McLaughlin} \&
  {Pudritz}}{1996}]{1996ApJ...457..578M}
{McLaughlin} D.~E.,  {Pudritz} R.~E.,  1996, \apj, 457, 578

\bibitem[\protect\citeauthoryear{{Mengel}, {Lehnert}, {Thatte} \&
  {Genzel}}{{Mengel} et~al.}{2005}]{2005A&A...443...41M}
{Mengel} S.,  {Lehnert} M.~D.,  {Thatte} N.,    {Genzel} R.,  2005, \aap, 443,
  41

\bibitem[\protect\citeauthoryear{{Metz}, {Cooper}, {Guerrero}, {Chu}, {Chen} \&
  {Gruendl}}{{Metz} et~al.}{2004}]{2004ApJ...605..725M}
{Metz} J.~M.,  {Cooper} R.~L.,  {Guerrero} M.~A.,  {Chu} Y.-H.,  {Chen}
  C.-H.~R.,    {Gruendl} R.~A.,  2004, \apj, 605, 725

\bibitem[\protect\citeauthoryear{{Meurer}}{{Meurer}}{1995}]{1995Natur.375..742%
M}
{Meurer} G.~R.,  1995, \nat, 375, 742

\bibitem[\protect\citeauthoryear{{Mirabel}, {Vigroux}, {Charmandaris},
  {Sauvage}, {Gallais}, {Tran}, {Cesarsky}, {Madden} \& {Duc}}{{Mirabel}
  et~al.}{1998}]{1998A&A...333L...1M}
{Mirabel} I.~F.,  {Vigroux} L.,  {Charmandaris} V.,  {Sauvage} M.,  {Gallais}
  P.,  {Tran} D.,  {Cesarsky} C.,  {Madden} S.~C.,    {Duc} P.-A.,  1998, \aap,
  333, L1

\bibitem[\protect\citeauthoryear{{Murgia}, {Crapsi}, {Moscadelli} \&
  {Gregorini}}{{Murgia} et~al.}{2002}]{2002A&A...385..412M}
{Murgia} M.,  {Crapsi} A.,  {Moscadelli} L.,    {Gregorini} L.,  2002, \aap,
  385, 412

\bibitem[\protect\citeauthoryear{{Neff} \& {Ulvestad}}{{Neff} \&
  {Ulvestad}}{2000}]{2000AJ....120..670N}
{Neff} S.~G.,  {Ulvestad} J.~S.,  2000, \aj, 120, 670

\bibitem[\protect\citeauthoryear{{Parmentier} \& {Gilmore}}{{Parmentier} \&
  {Gilmore}}{2005}]{2005MNRAS.363..326P}
{Parmentier} G.,  {Gilmore} G.,  2005, \mnras, 363, 326

\bibitem[\protect\citeauthoryear{{Parmentier} \& {Gilmore}}{{Parmentier} \&
  {Gilmore}}{2006}]{parmentier06}
{Parmentier} G.,  {Gilmore} G.,  2006, submitted to \mnras

\bibitem[\protect\citeauthoryear{{Patil} \& {Rao}}{{Patil} \&
  {Rao}}{1978}]{patil78}
{Patil} G.~P.,  {Rao} C.~R.,  1978, Biometrics, 34, 179

\bibitem[\protect\citeauthoryear{{Rathborne}, {Jackson} \& {Simon}}{{Rathborne}
  et~al.}{2006}]{2006ApJ...641..389R}
{Rathborne} J.~M.,  {Jackson} J.~M.,    {Simon} R.,  2006, \apj, 641, 389

\bibitem[\protect\citeauthoryear{{Rich}, {Shara} \& {Zurek}}{{Rich}
  et~al.}{2001}]{2001AJ....122..842R}
{Rich} R.~M.,  {Shara} M.~M.,    {Zurek} D.,  2001, \aj, 122, 842

\bibitem[\protect\citeauthoryear{{Rosolowsky}}{{Rosolowsky}}{2005}]{2005PASP..%
117.1403R}
{Rosolowsky} E.,  2005, \pasp, 117, 1403

\bibitem[\protect\citeauthoryear{{Sandage} \& {Tammann}}{{Sandage} \&
  {Tammann}}{1995}]{1995ApJ...446....1S}
{Sandage} A.,  {Tammann} G.~A.,  1995, \apj, 446, 1

\bibitem[\protect\citeauthoryear{{Saviane}, {Hibbard} \& {Rich}}{{Saviane}
  et~al.}{2004}]{2004AJ....127..660S}
{Saviane} I.,  {Hibbard} J.~E.,    {Rich} R.~M.,  2004, \aj, 127, 660

\bibitem[\protect\citeauthoryear{{Schlegel}, {Finkbeiner} \&
  {Davis}}{{Schlegel} et~al.}{1998}]{1998ApJ...500..525S}
{Schlegel} D.~J.,  {Finkbeiner} D.~P.,    {Davis} M.,  1998, \apj, 500, 525

\bibitem[\protect\citeauthoryear{{Schmidt}}{{Schmidt}}{1959}]{1959ApJ...129..2%
43S}
{Schmidt} M.,  1959, \apj, 129, 243

\bibitem[\protect\citeauthoryear{{Schulz}, {Fritze-v.~Alvensleben},
  {M{\"o}ller} \& {Fricke}}{{Schulz} et~al.}{2002}]{2002A&A...392....1S}
{Schulz} J.,  {Fritze-v.~Alvensleben} U.,  {M{\"o}ller} C.~S.,    {Fricke}
  K.~J.,  2002, \aap, 392, 1

\bibitem[\protect\citeauthoryear{{Schweizer}}{{Schweizer}}{2002}]{2002IAUS..20%
7..630S}
{Schweizer} F.,  2002, in {Geisler} D.,  {Grebel} E.~K.,   {Minniti} D.,  eds,
  IAU Symposium {Evolution of Globular Clusters Formed in Mergers}.
p.~630

\bibitem[\protect\citeauthoryear{{Schweizer}}{{Schweizer}}{2003}]{2003IAUJD..1%
1E..44S}
{Schweizer} F.,  2003, Dynamics and Evolution of Dense Stellar Systems, 25th
  meeting of the IAU, Joint Discussion 11, 18 July 2003, Sydney, Australia, 11

\bibitem[\protect\citeauthoryear{{Schweizer} \& {Seitzer}}{{Schweizer} \&
  {Seitzer}}{1998}]{1998AJ....116.2206S}
{Schweizer} F.,  {Seitzer} P.,  1998, \aj, 116, 2206

\bibitem[\protect\citeauthoryear{{Solomon}, {Downes} \& {Radford}}{{Solomon}
  et~al.}{1992}]{1992ApJ...387L..55S}
{Solomon} P.~M.,  {Downes} D.,    {Radford} S.~J.~E.,  1992, \apjl, 387, L55

\bibitem[\protect\citeauthoryear{{Solomon}, {Rivolo}, {Barrett} \&
  {Yahil}}{{Solomon} et~al.}{1987}]{1987ApJ...319..730S}
{Solomon} P.~M.,  {Rivolo} A.~R.,  {Barrett} J.,    {Yahil} A.,  1987, \apj,
  319, 730

\bibitem[\protect\citeauthoryear{{van den Bergh} \& {Lafontaine}}{{van den
  Bergh} \& {Lafontaine}}{1984}]{1984AJ.....89.1822V}
{van den Bergh} S.,  {Lafontaine} A.,  1984, \aj, 89, 1822

\bibitem[\protect\citeauthoryear{{Vesperini}}{{Vesperini}}{2000}]{2000MNRAS.31%
8..841V}
{Vesperini} E.,  2000, \mnras, 318, 841

\bibitem[\protect\citeauthoryear{{Vesperini}}{{Vesperini}}{2001}]{2001MNRAS.32%
2..247V}
{Vesperini} E.,  2001, \mnras, 322, 247

\bibitem[\protect\citeauthoryear{{Wang}, {Fazio}, {Ashby}, {Huang}, {Pahre},
  {Smith}, {Willner}, {Forrest}, {Pipher} \& {Surace}}{{Wang}
  et~al.}{2004}]{2004ApJS..154..193W}
{Wang} Z.,  {Fazio} G.~G.,  {Ashby} M.~L.~N.,  {Huang} J.-S.,  {Pahre} M.~A.,
  {Smith} H.~A.,  {Willner} S.~P.,  {Forrest} W.~J.,  {Pipher} J.~L.,
  {Surace} J.~A.,  2004, \apjs, 154, 193

\bibitem[\protect\citeauthoryear{{West}, {C{\^o}t{\'e}}, {Marzke} \&
  {Jord{\'a}n}}{{West} et~al.}{2004}]{2004Natur.427...31W}
{West} M.~J.,  {C{\^o}t{\'e}} P.,  {Marzke} R.~O.,    {Jord{\'a}n} A.,  2004,
  \nat, 427, 31

\bibitem[\protect\citeauthoryear{{Whitmore}}{{Whitmore}}{2004}]{2004ASPC..322.%
.419W}
{Whitmore} B.~C.,  2004, in {Lamers} H.~J.~G.~L.~M.,  {Smith} L.~J.,   {Nota}
  A.,  eds, ASP Conf. Ser. 322: The Formation and Evolution of Massive Young
  Star Clusters {Survival Rates and Consequences}.
pp 419--+

\bibitem[\protect\citeauthoryear{{Whitmore}, {Chandar} \& {Fall}}{{Whitmore}
  et~al.}{2006}]{2006astro.ph.11055W}
{Whitmore} B.~C.,  {Chandar} R.,    {Fall} S.~M.,  2006, ArXiv Astrophysics
  e-prints

\bibitem[\protect\citeauthoryear{{Whitmore}, {Gilmore}, {Leitherer}, {Fall},
  {Chandar}, {Blair}, {Schweizer}, {Zhang} \& {Miller}}{{Whitmore}
  et~al.}{2005}]{2005AJ....130.2104W}
{Whitmore} B.~C.,  {Gilmore} D.,  {Leitherer} C.,  {Fall} S.~M.,  {Chandar} R.,
   {Blair} W.~P.,  {Schweizer} F.,  {Zhang} Q.,    {Miller} B.~W.,  2005, \aj,
  130, 2104

\bibitem[\protect\citeauthoryear{{Whitmore} \& {Schweizer}}{{Whitmore} \&
  {Schweizer}}{1995}]{1995AJ....109..960W}
{Whitmore} B.~C.,  {Schweizer} F.,  1995, \aj, 109, 960

\bibitem[\protect\citeauthoryear{{Whitmore}, {Schweizer}, {Leitherer}, {Borne}
  \& {Robert}}{{Whitmore} et~al.}{1993}]{1993AJ....106.1354W}
{Whitmore} B.~C.,  {Schweizer} F.,  {Leitherer} C.,  {Borne} K.,    {Robert}
  C.,  1993, \aj, 106, 1354

\bibitem[\protect\citeauthoryear{{Whitmore} \& {Zhang}}{{Whitmore} \&
  {Zhang}}{2002}]{2002AJ....124.1418W}
{Whitmore} B.~C.,  {Zhang} Q.,  2002, \aj, 124, 1418

\bibitem[\protect\citeauthoryear{{Whitmore}, {Zhang}, {Leitherer}, {Fall},
  {Schweizer} \& {Miller}}{{Whitmore} et~al.}{1999}]{1999AJ....118.1551W}
{Whitmore} B.~C.,  {Zhang} Q.,  {Leitherer} C.,  {Fall} S.~M.,  {Schweizer} F.,
     {Miller} B.~W.,  1999, \aj, 118, 1551

\bibitem[\protect\citeauthoryear{{Wilson}, {Scoville}, {Madden} \&
  {Charmandaris}}{{Wilson} et~al.}{2003}]{2003ApJ...599.1049W}
{Wilson} C.~D.,  {Scoville} N.,  {Madden} S.~C.,    {Charmandaris} V.,  2003,
  \apj, 599, 1049

\bibitem[\protect\citeauthoryear{{Zepf}, {Ashman}, {English}, {Freeman} \&
  {Sharples}}{{Zepf} et~al.}{1999}]{1999AJ....118..752Z}
{Zepf} S.~E.,  {Ashman} K.~M.,  {English} J.,  {Freeman} K.~C.,    {Sharples}
  R.~M.,  1999, \aj, 118, 752

\bibitem[\protect\citeauthoryear{{Zhang} \& {Fall}}{{Zhang} \&
  {Fall}}{1999}]{1999ApJ...527L..81Z}
{Zhang} Q.,  {Fall} S.~M.,  1999, \apjl, 527, L81

\bibitem[\protect\citeauthoryear{{Zhang}, {Fall} \& {Whitmore}}{{Zhang}
  et~al.}{2001}]{2001ApJ...561..727Z}
{Zhang} Q.,  {Fall} S.~M.,    {Whitmore} B.~C.,  2001, \apj, 561, 727

\bibitem[\protect\citeauthoryear{{Zinchenko}, {Pirogov} \&
  {Toriseva}}{{Zinchenko} et~al.}{1998}]{1998A&AS..133..337Z}
{Zinchenko} I.,  {Pirogov} L.,    {Toriseva} M.,  1998, \aaps, 133, 337

\end{thebibliography}

\end{document}